\documentclass{pasj01}
\Received{$\langle$reception date$\rangle$}
\Accepted{$\langle$acception date$\rangle$}
\Published{$\langle$publication date$\rangle$}

\usepackage{natbib,bm}
\usepackage[varg]{txfonts}

\usepackage[pdftex]{graphicx,color}  

\newcommand{\beq}{\begin{equation}}
\newcommand{\eeq}{\end{equation}}
\newcommand{\beqn}{\begin{eqnarray}}
\newcommand{\eeqn}{\end{eqnarray}}
\newcommand{\la}{\lesssim}
\newcommand{\ga}{\gtrsim}
\newcommand{\eqref}[1]{(\ref{#1})}

\newcommand{\dfrac}[2]{{ \displaystyle\frac{#1}{#2}} }

\newcommand{\pfrac}[2]{ \biggl(\dfrac{#1}{#2}\biggr) }

\newcommand{\pd}{\partial}
 
\newcommand{\kB}{k_{\rm B}}

\newcommand{\eps}{\epsilon}

\begin{document}
\title{A global two-layer radiative transfer model for axisymmetric, shadowed protoplanetary disks}

\author{Satoshi \textsc{Okuzumi}\altaffilmark{1}}
\altaffiltext{1}{Department of Earth and Planetary Sciences, Tokyo Institute of Technology, Meguro, Tokyo 152-8551, Japan}
\email{okuzumi@eps.sci.titech.ac.jp}

\author{Takahiro \textsc{Ueda}\altaffilmark{2,3}}
\altaffiltext{2}{National Astronomical Observatory of Japan, Osawa 2-21-1, Mitaka, Tokyo 181-8588, Japan}
\altaffiltext{3}{Max-Planck Institute for Astronomy (MPIA), K\"{o}nigstuhl 17, D-69117 Heidelberg, Germany}

\author{Neal J. \textsc{Turner}\altaffilmark{4}}
\altaffiltext{4}{Jet Propulsion Laboratory, California Institute of Technology, Pasadena, CA 91109, USA}


\maketitle

\begin{abstract}
Understanding the thermal structure of protoplanetary disks is crucial for modeling planet formation and interpreting disk observations. We present a new two-layer radiative transfer model for computing the thermal structure of axisymmetric irradiated disks. Unlike the standard two-layer model, our model accounts for the radial as well as vertical transfer of the starlight reprocessed at the disk surface. The model thus allows us to compute the temperature below ``shadowed'' surfaces receiving no direct starlight. Thanks to the assumed axisymmetry, the reprocessed starlight flux is given in one-dimensional integral form that can be computed at a low cost. Furthermore, our model evolves the midplane temperature using a time-dependent energy equation and can therefore treat thermal instabilities. We apply our global two-layer model to disks with a planetary induced gap and confirm that the model reproduces the disks' temperature profiles obtained from more computationally expensive Monte Carlo radiative transfer calculations to an accuracy of less than 20\%. We also apply the model to study the long-term behavior of the thermal wave instability in irradiated disks. Being simple and computationally efficient, the global two-layer model will be suitable for studying the interplay between disks' thermal evolution and dust evolution.
\end{abstract}


\section{Introduction}
Protoplanetary disks' temperature structure governs many aspects of planet formation.
The radial compositional gradient of gas and solids caused by the radial temperature gradient may determine what planets forming at different positions are made of  \citep{Oberg11}. The location of the water snow line, where water ice sublimates and condenses, is particularly important in this context as it may constrain where rocky planets like the Earth  forms.  
Ice's sublimation, condensation, and sintering around the snow line can also cause a change in solid particles' stickiness \citep[e.g.,][]{Chokshi93,Dominik97,Wada09,SU17}\footnote{However, it is under debate whether silicates or water ice is stickier \citep[e.g.,][]{Kimura15,Gundlach15,Musiolik19}.} and a local pileup of silicates and ice \citep{Stevenson88,Cuzzi04,Saito11,S11a,Ida16,Schoonenberg17,Drazkowska17,Hyodo19}. These processes may not only affect planet formation directly but may also produce some observable features in disk thermal emission \citep{Banzatti15,ZBB15,Okuzumi16,Pinilla17}. 
Knowledge of disks' thermal structure is also necessary for interpreting scattered light's radial falloff and planet-launched spiral waves \citep[e.g.,][]{Isella18}, inferring turbulence from super-thermal linewidths \citep[e.g.,][]{Flaherty18,Flaherty20}, and  guiding planets' orbital migration \citep[e.g.,][]{Paardekooper10,Paardekooper11,Bitsch13}.

Despite its importance, our understanding of  disks' thermal structure is still limited. This is particularly true for the thermal structure deep inside the disks, where planet formation mainly occurs. The disk thermal structure well above the midplane has been well studied with observations of spectral energy distribution \citep[e.g.,][]{Kenyon87,DAlessio98,DAlessio99,Chiang97,Chiang01,Sierra20} and 
optically thick molecular emission lines \citep[e.g.,][]{Akiyama11,Rosenfeld13,Weaver18,Calahan21}. The temperature structure closer to the midplane can also be  constrained from intensity maps of marginally optically thin emission lines \citep{Zhang17} and channel maps of optically thick lines \citep{Dullemond20}, but there are only a few cases for which these approaches have been applied. The midplane temperature profile can also be inferred from multiwavelength imaging of dust continuum emission \citep{Kim19,Carrasco-Gonzalez19,Macias21}, but the inferred temperature depends on the assumed scattering properties of the opacity-dominating dust grains \citep[see][]{Carrasco-Gonzalez19}. 

The disk thermal structure is determined by stellar radiation, internal heating associated with disk accretion, and radiative cooling by gas and dust. In the classical framework of the viscous accretion model \citep{Lynden-Bell74}, accretion heating is the dominant heating mechanism in the inner few au midplane region, thus controlling the location of the snow line \citep[e.g.,][]{Davis05,Garaud07,Oka11,Bitsch15}. However, recent studies of the magnetohydrodynamics of weakly ionized protoplanetary disks have shown that  accretion heating favors the disk surface because the midplane region is too poorly ionized to sustain a large electric current \citep{Hirose11,Mori19}. These models predict that the magnetohydrodynamic accretion heating can  raise the midplane temperature only when the disk opacity is high enough \citep{Bethune20,Mori21}. This suggests that stellar irradiation rather than internal heating may determine the location of the water snow line.
 
The study of the thermal structure of passively irradiated protoplanetary disks has a long history \citep[e.g.,][]{Kusaka70,Calvet91,Chiang97,DAlessio98,Dullemond01,Dullemond02}, but its complete understanding has yet to be established because of two complications. The first complication is that the disk thermal structure entirely depends on the global distribution of small dust grains that absorb and reprocess stellar radiation. Because starlight grazes the disk surface at a small angle (typically $\sim 0.01$--0.1 radians), even a small perturbation on the surface can yield a shadowed region that does not directly receive stellar radiation \citep{Dullemond01,Dullemond04a,Dullemond04b}. Shadowed regions can have significantly low temperatures and hence can have gas and dust compositions that differ greatly from their surroundings \citep{Ohno21}.  However, to accurately compute the temperature of shadowed regions, one must account for the radial radiative transfer of reprocessed starlight  \citep[e.g.,][]{Jang-Condell03,Dullemond04a,Turner12,Jang-Condell12,Ueda17}.
The growth, settling, and radial migration of small dust grains should also be taken into account as these processes may change the distribution of the shadows.
 
The second complication concerns the instabilities of disks' thermal structure. \citet{Dullemond00} and \citet{Watanabe08}  showed that optically thick disks irradiated by the central star are unstable to self-shadowing in the limits of short and long thermal relaxation timescales, respectively  \citep[for more recent studies, see][]{Siebenmorgen12,Ueda21,Wu21,Pavlyuchenkov22}. This instability, which we call the thermal wave instability after \citet{Watanabe08}\footnote{We avoid calling this the irradiation instability \citep{Wu21} because this term was previously used for a different disk instability \citep{Fung14}.}, is intrinsically related to the starlight's small grazing angle mentioned above: even a small hill on the irradiation surface can cast a shadow. The hill's starlit inner side warms and expands, while its shadowed outer side cools and contracts, so that the hill propagates towards the star (see figure 1 of \citealt{Ueda21} and \citealt{Wu21} for a cartoon describing the mechanism of the instability). 
The surface waves generated by the instability produce the interior temperature's fluctuations that propagate inward on a timescale comparable to the local thermal relaxation timescale at the midplane. 
The instability is potentially relevant to planet formation because it can cause snow lines to oscillate radially \citep{Ueda21} and may even produce pressure maxima collecting pebble-sized particles \citep{Watanabe08}. 
More recently, \citet{Owen20} has identified a distinct type of thermal instability that results from an abrupt change in the opacity across a snow line.
However, fully understanding these thermal instabilities requires a radiative transfer model that can treat shadowing and does not assume radiative equilibrium.
Lacking a simple model that fulfills these requirements, the roles of the thermal instabilities in planet formation have not been elucidated so far.

In this study, we present a simple radiative transfer model that can treat shadowed protoplanetary disks. Our model is a generalization of the well-known two-layer model \citep{Kusaka70,Chiang97,Chiang01}, which computes the interior temperature of an optically thick disk by considering the starlight reprocessed on the disk surface as the heating source. Being extremely simple and moderately accurate \citep[see][]{Dullemond04a}, the two-layer model has been widely used to model the radial temperature distribution of irradiated protoplanetary disks. However, the conventional two-layer model only considers the vertical transfer of the reprocessed starlight and hence is unable to compute temperatures just below shadowed surfaces. Our new model, which we call the global two-layer model, resolves this issue by accounting for the radial transfer of the reprocessed starlight. Our model is limited to axisymmetric disks, but is much simpler than exising radiative transfer models that can treat shadows, including those based on the Monte Carlo approach \citep{Dullemond12,Turner12} and those considering three-dimensional perturbations on the disk surface \citep{Jang-Condell03,Jang-Condell04,Jang-Condell08,Jang-Condell09,Jang-Condell12}. Therefore, our model will be suitable for coupled simulations of dust evolution and  disk thermal evolution.
Furthermore, our model does not rely on radiative equilibrium and hence can treat disks' thermal instabilities. 
The aims of this paper are to formulate and the global two-layer model and demonstrate its applicability to shadowed disks.

The structure of this paper is as follows.
Section~\ref{sec:classical} reviews the standard local two-layer model relying on the plane-parallel approximation and highlights its limitation.
Section~\ref{sec:model} describes our new two-layer model, and section~\ref{sec:validation} tests the model by comparing it with Monte Carlo radiative transfer calculations for gapped disks.
In section~\ref{sec:twi}, we use the model to simulate long-term evolution of the thermal wave instability. Section~\ref{sec:summary} presents a summary and future directions.

\section{Local two-layer model}\label{sec:classical}
\begin{figure*}[t]
\begin{center}
\includegraphics[width=16cm, bb=0 0 850 250]{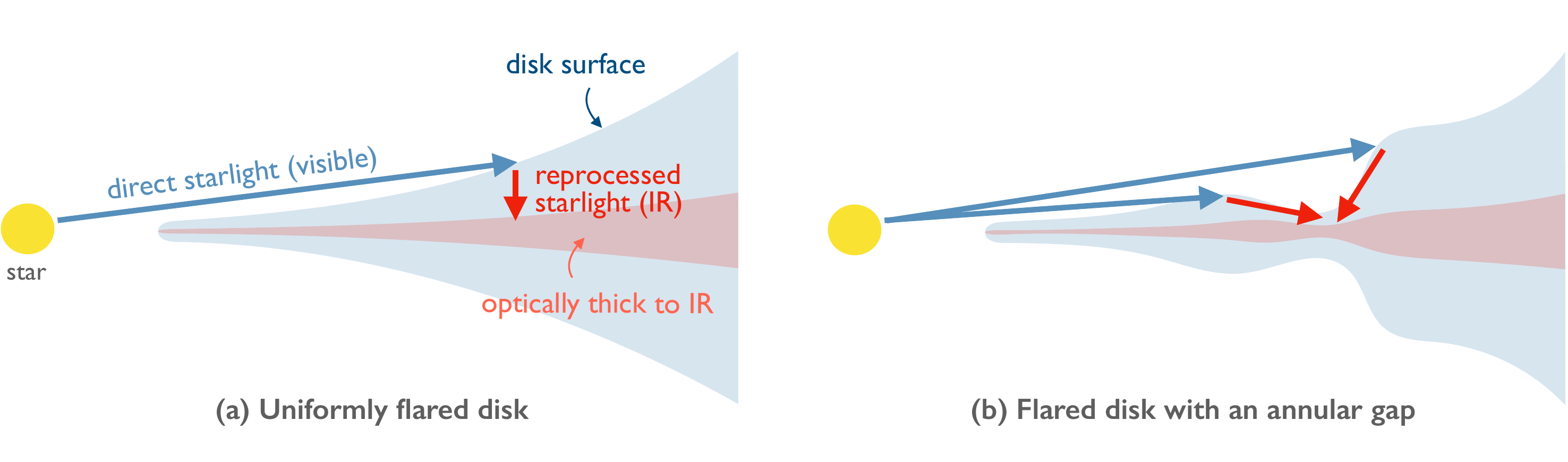}
\end{center}
\caption{Schematic showing the transfer of stellar radiation in a uniformly-flared disk (panel a) and in a flared disk with an annular gap, i.e., a dip in the radial surface density profile (panel b). The disk surface here is the location where vertical optical depth at visible wavelengths reaches $\sim H/r$, {where $H$ is the disk's scale height. For flared disks, this surface well approximates the true irradiation surface where the optical depth along starlight rays reaches unity (section~\ref{sec:classical})}. 
{Flared disk surfaces receive direct (visible) starlight and reprocess it into infrared (IR) radiation.} The conventional, local two-layer model computes the disk's interior temperature using the vertical transfer of the reprocessed starlight (panel a).
The local two-layer model does not apply to deep gaps that receive no direct starlight. To obtain the temperature in such a shadowed gap, one must consider the radial transfer of reprocessed starlight (panel b).
}
\label{fig:1}
\end{figure*}

We begin by reviewing the conventional local two-layer radiative transfer model for passively irradiated disks \citep[][{see also \citealt{Krugel08,Armitage10} for textbook descriptions}]{Chiang97,Chiang01}, with a particular emphasis on its key assumptions and limitations. The aim here is to explain why the local two-layer model is inapplicable to disks with shadows.

The local two-layer model assumes that (1) the disk is locally plane-parallel, i.e., the disk structure changes radially on a scale much longer than the disk vertical thickness, and that (2) the disk is flared, so that every portion of the disk surface receives direct starlight {(see figure~\ref{fig:1} for a schematic)}. Here, the disk surface refers to where the vertical optical depth at optical wavelengths is {$\sim H/r$, where $H$ is the disk's scale height.} For flared disks, this surface well approximates the true irradiation surface where the optical depth to the grazing starlight reaches unity  {because the radial optical depth is $\sim r/H$ times larger than the vertical depth \citep[in other words, the starlight grazing angle for flared disks is comparable to $H/r$; see, e.g., ][]{Chiang97}}. 

\begin{figure}[t]
\begin{center}
\includegraphics[width=8cm, bb=0 0 260 187]{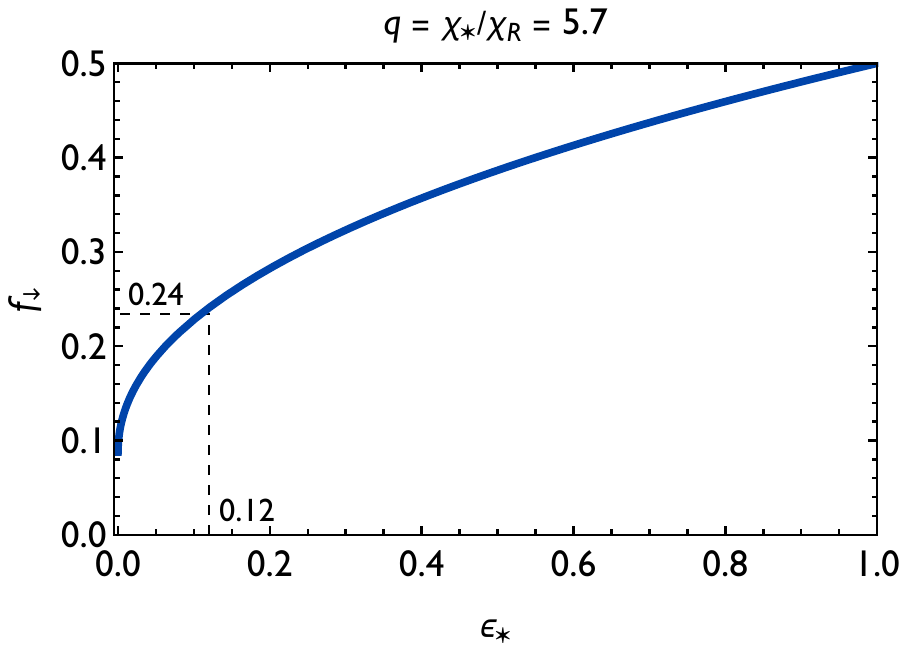}
\end{center}
\caption{Fraction $f_\downarrow$ of the stellar radiation reprocessed downward as a function of $\eps_*$ for the case $q = \chi_*/\chi_{\rm R} = 5.7$ considered in section~\ref{sec:validation}. Here, $\eps_*$ is unity minus the grains' starlight albedo. {A value of $\eps_*< 1$ yields $f_\downarrow < 1/2$; for instance,  $\eps_* = 0.12$ gives $f_\downarrow = 0.24$.}   
}
\label{fig:f_down}
\end{figure}

With the plane-parallel and flared assumptions, the vertical downward flux of the reprocessed (infrared) radiation from the irradiation surface toward the disk interior is given by 
\beq
F_{\rm rep, \downarrow} = f_\downarrow \mu_* F_*, 
\label{eq:Fs_planeparallel}
\eeq
where $F_*$ is the magnitude of the direct starlight flux,
$\mu_*$ is the sine of the grazing angle between the starlight and disk surface, and  $f_\downarrow$ a dimensionless coefficient that depends on the single-scattering albedo of the {grains receiving the starlight}.
In equation~\eqref{eq:Fs_planeparallel}, the product $\mu_* F_*$ represents the flux of the stellar radiation incident on the irradiation surface, while the prefactor $f_\downarrow$ represents the fraction of the starlight flux reprocessed downward. 
A purely absorbing atmosphere gives $f_\downarrow = 1/2$, meaning that it 
reprocesses half of the incident stellar radiation energy into downward thermal radiation \citep{Chiang97}. 
For a plane-parallel atmosphere of nonzero albedo, one has (see appendix~\ref{appendixA} for a derivation)
\beq
f_\downarrow = 
\frac{\eps_*}{2} + 
\left(\sqrt{3\eps_*}+\frac{3}{2q}\right)\frac{1-\eps_*}{3+2\sqrt{3\eps_*}},
\label{eq:f_down}
\eeq
where
\beq
\eps_* = \frac{\kappa_{*} }{\chi_{*} }
\label{eq:epsstar}
\eeq
is the ratio between the Planck-mean absorption and extinction opacities $\kappa_{*}$ and $\chi_{*}$ evaluated at the stellar surface temperature $T_*$, and 
\beq
q = \frac{\chi_*}{\chi_{\rm R}}
\eeq
is the ratio between $\chi_*$ and the Rosseland mean extinction opacity for the disk's own thermal radiation, $\chi_{\rm R}$ {(see equations~\eqref{eq:kappa_star}, \eqref{eq:chi_star}, and \eqref{eq:chiR} for the definition of the mean opacities)}. {The ratio $\eps_*$ is related to the grains' single-scattering albedo  $\omega_*$ for the starlight as $\omega_* = 1-\eps_*$.}
The fraction $f_\downarrow$ decreases monotonically with decreasing $\eps_*$, reflecting the fact that multiple scattering by the gas and dust particles enhances the backscattering of the starlight.
{Figure~\ref{fig:f_down} illustrates $f_\downarrow$ as a function of $\eps_*$ for the particular case of $q=5.7$ considered in section~\ref{sec:validation}. }

Assuming that the disk interior is optically thick to infrared radiation and is also in radiative equilibrium, the downward infrared flux given by equation~\eqref{eq:Fs_planeparallel} balances the upward thermal flux from the interior, $F_i = \sigma_{\rm SB} T^4$, where $T$ is the interior temperature and $\sigma_{\rm SB}$ is the Stefan--Boltzmann constant. This balance yields 
\beq
T(r) = \pfrac{f_\downarrow\mu_*  F_*}{\sigma_{\rm SB}}^{1/4} .
\label{eq:T_parallel}
\eeq
For a central star of radius $R_*$ and surface temperature $T_*$, $F_*$ can be written as 
\beq
F_* = \frac{\sigma_{\rm SB} R_*^2 T_*^4}{r^2+z_s(r)^2} \approx \frac{\sigma_{\rm SB} R_*^2 T_*^4}{r^2},
\label{eq:Fstar}
\eeq
where $r$ is the cylindrical distance from the central star and $z_s(r)$ is the height of the irradiation surface at $r$.  
The final expression assumes $z_s(r)^2/r^2 \ll 1$, which holds in typical protoplanetary disks. 
Substituting equation~\eqref{eq:Fstar} into \eqref{eq:T_parallel} yields
\beq
T(r) = (f_\downarrow\mu_*)^{1/4} \pfrac{R_*}{r}^{1/2} T_*.
\label{eq:T_parallel_2}
\eeq
For purely absorbing atmospheres ($f_\downarrow = 1/2$), equation~\eqref{eq:T_parallel_2} reduces to the well-known expression for the disk interior temperature in the conventional two-layer model neglecting scattering  (e.g., equation~(4) of \citealt{Dullemond01}).

It is important to note  that every quantity in the right hand side of equation~\eqref{eq:T_parallel}, or equivalently of equation~\eqref{eq:T_parallel_2}, is evaluated locally. In other words, the conventional two-layer model only accounts for the vertical transfer of reprocessed starlight as schematically shown in figure~\ref{fig:1}(a). 

However, such a model is incompatible with disks with a shadow. To illustrate this, we consider a disk with an annular gap, i.e., a local dip in the radial surface density profile (figure~\ref{fig:1}(b)).
For the moment, we continue to refer to the location where the vertical visible optically depth is $\sim H/r$ as the disk surface. As shown in figure~\ref{fig:1}(b), this surface has a valley in the gap. We now assume that the valley is so deep that it  falls into the shadow cast by the region inward of the gap. The problem is that such a valley receives no direct starlight and therefore provides no reprocessed radiation toward the midplane. Hence, one cannot use equation~\eqref{eq:T_parallel} or \eqref{eq:T_parallel_2} to compute the midplane temperature in the shadowed gap. This example clearly indicates that the radial transfer of reprocessed radiation must be taken into account to compute the temperature in shadowed regions (see figure~\ref{fig:1}(b)).

\section{Global two-layer model}\label{sec:model}
In this section, we present our global two-layer model that takes into account the radial as well as vertical transfer of reprocessed starlight. As in the standard two-layer model, we consider the irradiation surface reprocessing direct (visible) starlight into thermal (infrared) radiation and the disk interior heated by the reprocessed starlight (sections~\ref{sec:irradiation} and \ref{sec:photosphere}). We relax the two fundamental assumptions of the standard two-layer model by splitting the irradiation surface into concentric rings and calculate the two-dimensional transfer of the downward emission from individual rings (section~\ref{sec:reprocessed}). 
Our current model assumes vertical hydrostatic equilibrium (section~\ref{sec:vertical}) but does not assume radiative equilibrium (section~\ref{sec:evolution}).  We also present numerical implementation of the model (section~\ref{sec:numerics}).

Our global two-layer model is largely inspired by the radiative transfer model of \citet{Jang-Condell03,Jang-Condell04} and \citet{Jang-Condell08}, which splits the irradiation surface into small elements and computes three-dimensional radiative transfer of the reprocessed starlight from each element. Compared to this previous model, our model is limited to axisymmetric disks. However, this assumed symmetry allows us to  greatly simplify the expression of the reprocessed starlight flux  (for details, see section~\ref{sec:reprocessed}).

In the following, we employ the cylindrical coordinate system $(r,\phi,z)$ centered on the central star.
Because we assume axisymmetry, every  quantity is independent of $\phi$. 

\subsection{The irradiation surface}\label{sec:irradiation}
To properly treat shadows, we do not use an approximate irradiation surface defined in terms of the vertical optical depth as used in figure~\ref{fig:1}. 
We employ the exact irradiation surface defined as where the optical depth to the direct starlight is unity \citep{DAlessio01}. 
At position $(r,z)$ in a disk, the Planck-mean optical depth to the direct starlight is given by
\beq
\tau_{*} = \int_{\rm starlight} \chi_{*} \rho \,ds, 
\label{eq:taustar}
\eeq
where $\chi_*$ is the Planck-mean extinction opacity per gas mass for the starlight, $\rho$ is the gas mass density, and $s$ denotes the path length of the stellar ray from the star to the position considered.
Thus, the height of the irradiation surface, $z_s (r)$, at arbitrary cylindrical radius $r$ 
is defined by the relation $\tau_*(r,z_s(r))=1$.
The disk's vertical structure is specified in section~\ref{sec:vertical}.

\begin{figure}
\begin{center}
\includegraphics[width=8cm, bb=0 0 400 350]{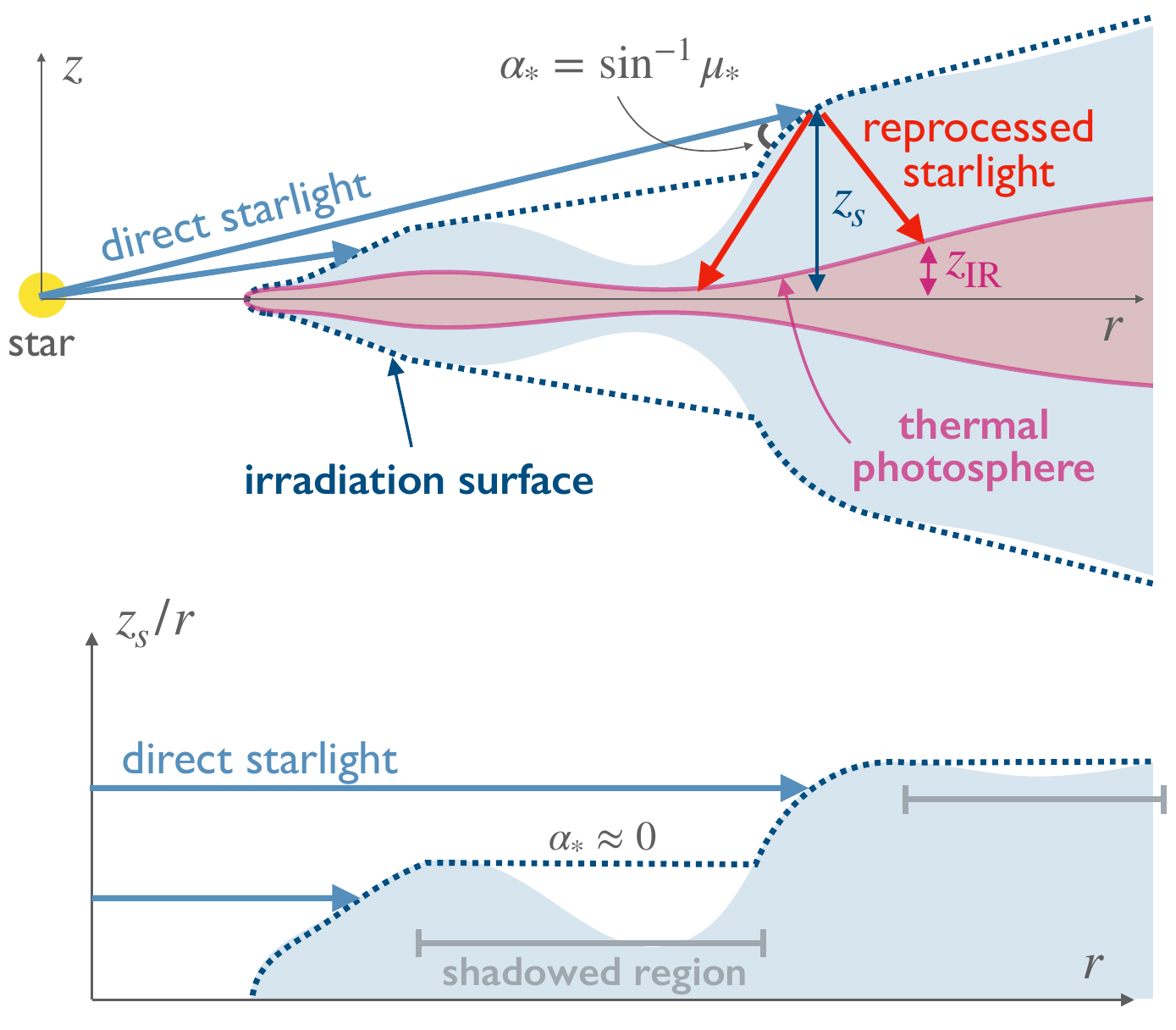}
\end{center}
\caption{Upper panel: schematic showing a disk's irradiation surface (dotted curve; see section~\ref{sec:irradiation}) and thermal (infrared) photophere (thin solid curve; see section~\ref{sec:photosphere}). The quantities $\alpha_*$, $z_s$, and $z_{\rm IR}$ stand for the starlight grazing angle, irradiation surface's height, and thermal photosphere's height, respectively. Lower panel: the irradiation surface as seen in the $r$--$z/r$ plane. In this plane, stellar rays of different propagation angles are represented by horizontal lines. A shadowed region refers to where $\alpha_*$ vanishes. As depicted in figure \ref{fig:1}(b), this region receives no direct stellar radiation but is heated by the reprocessed starlight that propages both radially and vertically.}

\label{fig:irr}
\end{figure}
At large radial distances where the central star can be seen as a point source, stellar rays travel along lines of approximately constant $z/r$. For such regions, it is useful to map the irradiation surface in the $r$--$z/r$ plane. 
Figure~\ref{fig:irr} schematically shows the irradiation surface of the gapped disk shown in figure~\ref{fig:1}(b). 
As already mentioned in section~\ref{sec:classical}, the irradiation surface outside the shadowed gap approximately matches the surface at which the vertical visual optical depth is $\sim H/r$. 
Inside the shadowed gap, the irradiation surface as seen in the $r$--$z/r$ plane is represented by a horizontal line. Below this line, the material inward of the gap blocks the direct starlight (see the lower panel of figure~\ref{fig:irr}).

For given $z_s(r)$, one can calculate $\mu_* = \sin \alpha_*$ as \citep{Kusaka70,Ruden91}
\beq
\mu_* = \sin^{-1}\pfrac{4R_*}{3\pi r} 
+ \tan^{-1}\left(\frac{z_s}{r}\frac{d\ln z_s}{d\ln r}\right) 
- \tan^{-1}\left(\frac{z_s}{r}\right) .
\label{eq:mustar}
\eeq
In the right-hand side of equation~\eqref{eq:mustar}, the first term accounts for the finite stellar size, while the sum of the second and third terms is related to the grazing angle of the ray from the central point source.
At large radial distances where the first term is negligible, $\mu_*$ vanishes in shadowed regions with radially constant  $z_s/r$ (see the lower panel of figure~\ref{fig:irr}).

\subsection{The thermal photosphere}\label{sec:photosphere}
{We define the thermal photosphere as} the surface on which the optical depth to reprocessed starlight {from the irradiation surface} reaches unity {(see figure~\ref{fig:irr})}. Strictly speaking, the location of the thermal photosphere depends on the incident angle of the reprocessed radiation. To avoid this complexity, we approximate the thermal photosphere by the surface where the {\it vertical} optical depth for downward infrared radiation exceeds $1/2$. This choice is based on the fact that the angle-averaged optical depth of an optically thin disk to isotropic radiation is twice the vertical optical depth \citep{Nakamoto94}.
Writing the Planck-mean extinction opacity for infrared radiation as $\chi_{\rm P}$, the corresponding vertical optical depth is given by
\beq
\tau_{\rm P}(r,z) = \int^\infty_z \chi_{\rm P}(r,z')\rho(r,z')\, dz'. 
\label{eq:tauP}
\eeq
Using this, the height of the thermal photosphere, $z_{\rm IR}(r)$, at radial position $r$ can be approximately defined by the relation $\tau_{\rm P}(r,z_{\rm IR}(r)) = 1/2$. {The region well below the thermal photosphere is also optically thick to its own thermal emission.}

For disks with $\tau_{\rm P} < 1/2$ at the midplane, we set $z_{\rm IR} = 0$ for convenience. However, the midplane of such an optically thin disk is not a real thermal photosphere, absorbing only a small fraction of the incoming infrared radiation. The low absorptivity and emissivity of optically thin disks are taken into account in section~\ref{sec:evolution}.

\subsection{Flux of the downward reprocessed starlight}\label{sec:reprocessed}

\begin{figure*}
\begin{center}
\includegraphics[width=11cm, bb=0 0 440 230]{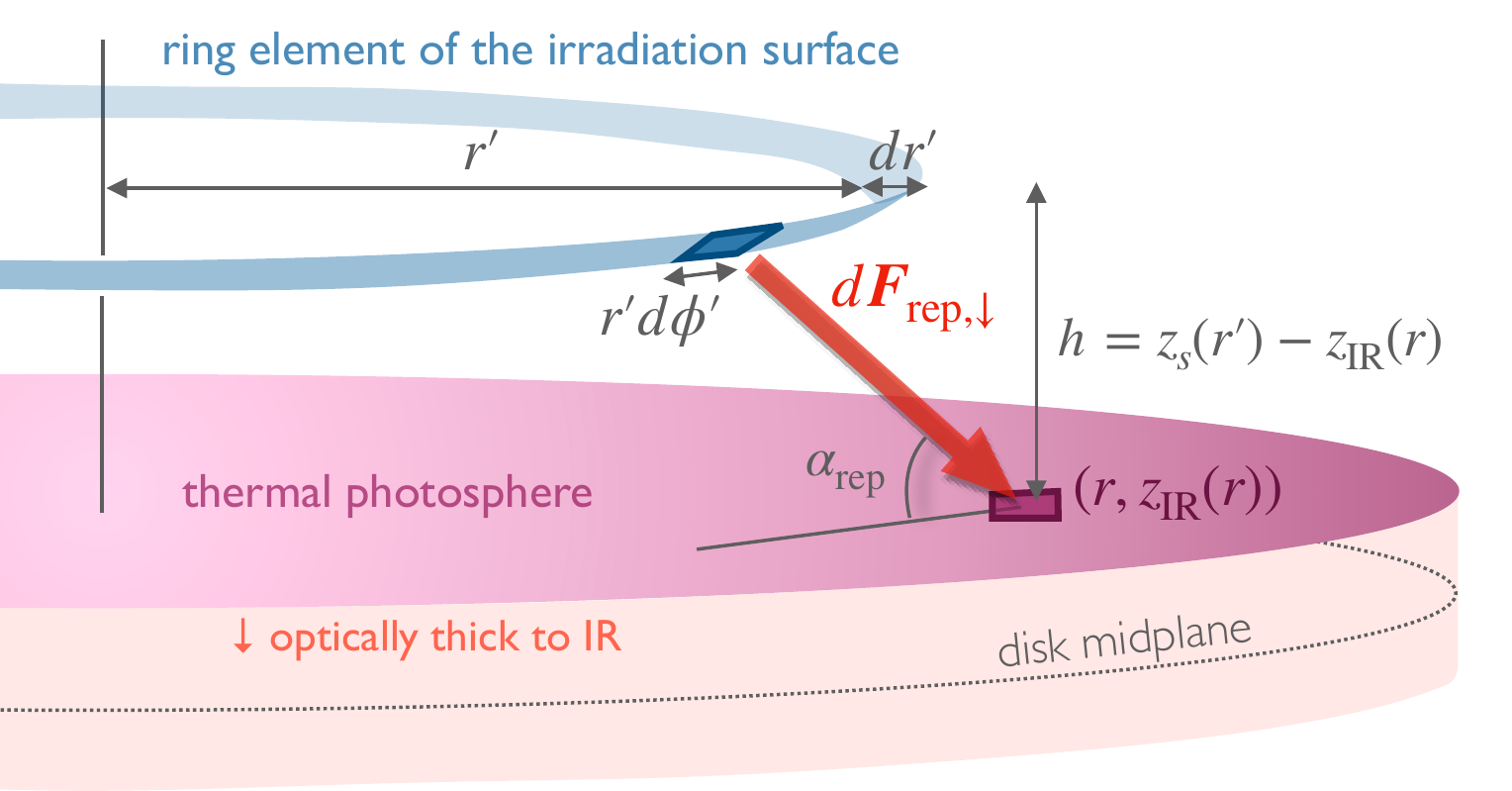}
\end{center}
\caption{Schematic showing how the global two-layer model 
computes the flux of the reprocessed starlight from the irradiation surface to the optically thick interior.
The irradiation surface is decomposed into thin ring elements of radius $r'$  lying at height $z = z_s(r')$
above the midplane. Each ring element is further divided into segments of azimuthal width $r'd\phi'$. The disk interior that is optically thick to infrared thermal radiation receives the reprocessed starlight on its surface (thermal photosphere) lying at height $z = z_{\rm IR}(r)$. The flux of the reprocessed starlight from each ring element to position $(r, z_{\rm IR}(r))$ is calculated by integrating the contributions $d{\bm F}_{\rm rep,\downarrow}$ from the constituting segments.
}
\label{fig:F_IR}
\end{figure*}
We split the irradiation surface into thin concentric ring elements of radius $r'$, radial extent ${dr'}$, and height $z_s(r')$  (figure~\ref{fig:F_IR}), Each ring emits infrared radiation with a luminosity proportional to the received stellar flux. 
To derive an analytic expression for the downward radiation flux from each ring element, we further divide the rings azimuthally into rectangular segments of azimuthal width $r'd\phi'$. The area of each segment is  $dS = \sqrt{dr'^2 + dz_s^2} \, r' d\phi' = \sqrt{1 + (dz_s/dr')^2}\, r'  dr' d\phi' $. 
The luminosity of the downward radiation from each segment is  
\beqn
dL_{\rm rep, \downarrow}  &=& f_\downarrow(r') \mu_* F_* dS 
\nonumber \\
&=& f_\downarrow \mu_* F_* \sqrt{1 + (dz_s/dr')^2}\, r' d r' d\phi'.
\eeqn

The radiation from each surface segment strikes the thermal photosphere. 
For simplicity, we here assume $dz_{\rm IR}/dr \ll 1$ and approximate the thermal photosphere as a plane locally parallel to the midplane. Because of this approximation, our model neglects {the shadowing of the reprocessed radiation field} on the thermal photosphere. 
For typical protoplanetary disks with $z_{\rm IR} \ll r$, the approximation made here can be justified unless  $z_{\rm IR}$ varies on a radial lengthscale $\ll r$. 
Because we assume axisymmetry, the azimuthal coordinate of an arbitrary segment of the thermal photosphere can be taken to be zero without loss of generality.
The angle between the reprocessed starlight and the thermal photosphere segment can then be written by 
\beq
\alpha_{\rm rep} = \sin^{-1}\pfrac{h}{\Delta},
\label{eq:alpha_IR}
\eeq
where 
\beq
h(r,r') \equiv z_s(r') - z_{\rm IR}(r)
\label{eq:h}
\eeq
is the height of the irradiation surface segment relative to the photosphere segment and 
\beq
\Delta \equiv \sqrt{(r-r' \cos\phi' )^2 + r'^2 \sin^2\phi' + h^2}
\label{eq:Delta}
\eeq 
is the distance between the two segments. 

Assuming that the downward reprocessed starlight is isotropic, and approximating the irradiation surface segment as a point source, the magnitude of the downward reprocessed starlight flux $d{\bm F}_{\rm rep, \downarrow}$ (see figure~\ref{fig:F_IR}) can be written as 

\beq
|d{\bm F}_{\rm rep, \downarrow}| = \frac{dL_{\rm rep, \downarrow}}{2\pi\Delta^2},
\eeq
where the factor $2\pi$ comes from the solid angle of the lower hemisphere.
Because the reprocessed starlight strikes at angle $\alpha_{\rm rep}$, 
its flux incident on the thermal photosphere is 
$\sin\alpha_{\rm rep}|d{\bm F}_{\rm rep, \downarrow}|$.
Integrating this over $\phi'$, we obtain 
\beqn
&& 
\int_{\phi'=0}^{\phi'=2\pi} \sin\alpha_{\rm rep}|d{\bm F}_{\rm rep, \downarrow}| 
\nonumber \\
&&= \frac{f_\downarrow \mu_* F_* h\sqrt{1 + (dz_s/dr')^2}\, r' dr' }{2\pi} \int_0^{2\pi} \frac{d\phi' }{\Delta^3}
\nonumber \\
&&=  
\frac{2f_\downarrow \mu_* F_* h \sqrt{1 + (dz_s/dr')^2}\, r' d r'\, E(k)} {\pi ( (r'-r)^2 + h^2) \sqrt{(r'+r)^2 + h^2} },
\label{eq:F_ring}
\eeqn
where  $E(k) = \int_0^{\pi/2} (1-k^2 \sin^2 x)^{1/2} dx$ 
is the complete elliptic integral of the second kind
with $k =2\sqrt{rr'/((r'+r)^2+h^2)} $.
Finally, by integrating equation~\eqref{eq:F_ring} over $r'$, we obtain the net downward vertical flux $F_{\rm rep, \downarrow}(r)$ of the reprocessed starlight from the entire irradiation surface to the disk interior at radius $r$,
\beqn
F_{\rm rep, \downarrow}(r) &=& 
\int_{\phi', r'} \sin\alpha_{\rm rep}|d{\bm F}_{\rm rep, \downarrow}| 
\nonumber \\
&=& \frac{2}{\pi} 
\int \frac{f_\downarrow \mu_* F_* h\sqrt{1 + (dz_s/dr')^2}\, r' E(k)} {\left( (r'-r)^2 +h^2\right) \sqrt{(r'+r)^2 +h^2} }dr'.
\nonumber \\
\label{eq:Fs}
\eeqn
Here, the integration is performed over regions of $\alpha_{\rm rep} > 0$ $(h > 0)$\footnote{The quantity $h$ can become negative because it compares $z_s$ and $z_{\rm IR}$ at different radial locations. For instance, $z_s$ at small $r'$ can be smaller than $z_{\rm IR}$ at large $r$.}. 
Note that $f_\downarrow$, $\mu_*$, $F_*$, and $h$ generally depend on $r'$ and hence should be inside the integral.

Equation~\eqref{eq:Fs} is a generalization of the reprocessed starlight flux in the local two-layer model, 
equation~\eqref{eq:Fs_planeparallel}.
To see this, we rewrite equation~\eqref{eq:Fs} as 
\beqn
F_{\rm rep, \downarrow}(r) 
= \int f_\downarrow(r') \mu_*(r')F_*(r') W(r,r')dr',
\label{eq:Fs_2}
\eeqn
where the function $W(r,r')$ is 
\beq
W(r,r') = \frac{2h\sqrt{1 + (dz_s/dr')^2}\, r' E(k)} 
{\pi \left( (r'-r)^2 +h^2\right) \sqrt{(r'+r)^2 +h^2} }
\eeq
for $h > 0$ and $W (r,r') = 0$ otherwise.
Equation~\eqref{eq:Fs_2} can be viewed as the radial average of equation~\eqref{eq:Fs_planeparallel} weighted by $W(r,r')$. 
As shown in appendix~\ref{appendixB}, equation~\eqref{eq:Fs_2} recovers equation~\eqref{eq:Fs_planeparallel} in the limit of small $h$.

\begin{figure}[t]
\begin{center}
\includegraphics[width=8cm, bb=0 0 288 201]{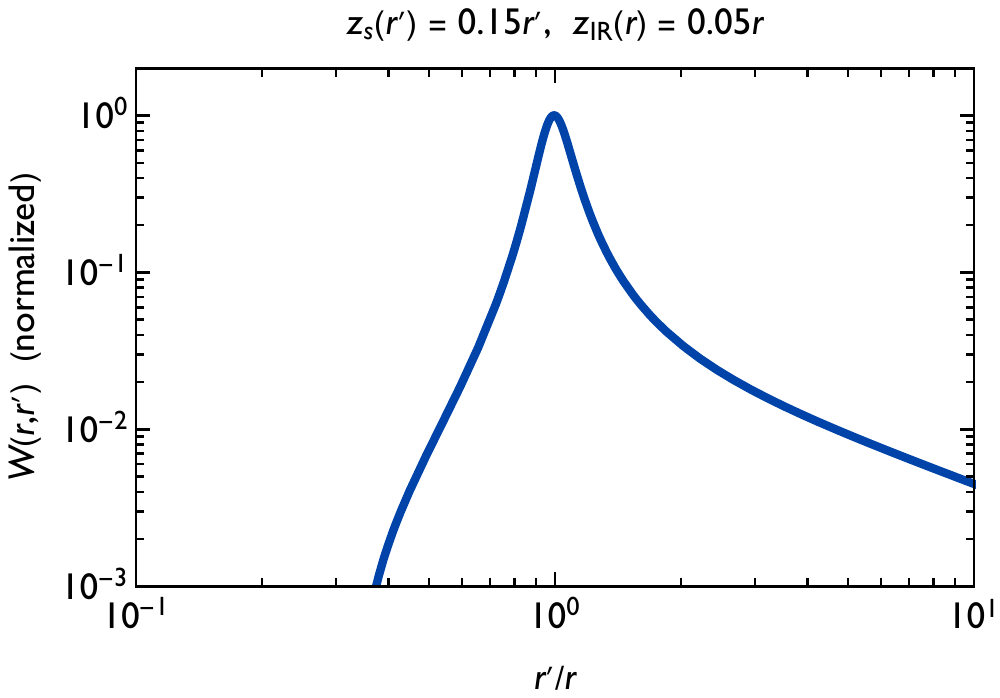}
\end{center}
\caption{Weighting function $W(r,r')$ in the radial integration in equation~\eqref{eq:Fs_2} as a function of $r'/r$ for $z_s =  0.15 r'$ and $z_{\rm IR} = 0.05r$, normalized such that $W=1$ at $r'=r$. 
}
\label{fig:W}
\end{figure}
The weighting function $W$ represents the radially nonlocal nature of the reprocessed radiation. 
As explained below, this function is  peaked around $r' = r$ and has a width of $\sim h$.
Because the elliptic integral $E$ is bounded in the narrow range of $1 \leq E \leq \pi/2$ and because $\sqrt{1 + (dz_s/dr')^2} \sim 1$, the profile of $W$ is essentially determined by the factor $hr'/[  ( (r'-r)^2 +h^2) \sqrt{(r'+r)^2 +h^2} ]$.
Since $h \ll r$, the profile around the peak is approximately Lorentzian $\propto 1/ ( (r-r')^2 + h^2 )$ with the half width at half maximum of $h$.
Figure~\ref{fig:W} illustrates the functional form of $W$ for $z_s(r') = 0.15 r'$ and $z_{\rm IR}(r) = 0.05r$, i.e., $h =  0.15 r' - 0.05r$. In this particular example, $W$ vanishes at $r'/r < 1/3$, at which $h$ becomes negative.

\subsection{Disk vertical structure}\label{sec:vertical}
So far, we have not specified the disk's vertical structure.
In the present study, we avoid detailed modeling of the vertical structure by approximating the disk as vertically isothermal and hydrostatic. 
These two approximations yield the vertical distribution of $\rho$ of the simple analytic form
\beq
\rho(r,z) = \frac{\Sigma(r)}{\sqrt{2\pi} H(r)}\exp\left(-\frac{z^2}{2H(r)^2}\right),
\label{eq:rho}
\eeq
where $\Sigma$ is the gas surface density and $H = c_s/\Omega_{\rm K}$ is the gas scale height, 
with $c_s$ and $\Omega_{\rm K}$ being the isothermal sound speed and Keplerian frequency, respectively.
The isothermal sound speed is related to the disk interior temperature as $c_s = \sqrt{\kB T/m_g}$, where $\kB$ is the Boltzmann constant and $m_g$ is the mean mass of the gas molecules.

We comment on the validity and limitations of the vertical isothermal and hydrostatic approximations. 
For passively irradiated disks in radiative equilibrium, the vertical isothermal approximation is valid well below the irradiation surface. However, this approximation breaks down across the irradiation surface, above which the temperature is higher because of direct stellar irradiation \citep{Calvet91,Chiang97,DAlessio98}. 
A more reasonable approximation would be to assign different temperatures to the interior ($z <z_s$) and warmer surface region ($z > z_s$) {as described in appendix B of \citet{Watanabe08}}. However, {such two-temperature modeling would result in a vertical density profile dependent on $z_s$, requiring} iterative determination of $z_s$ and $\rho(z)$. We defer this complication to future work. {Regarding the equilibrium disk structure, our vertical isothermal approximation should suffice to determine $z_s$ because the warmer but optically thin gas well above the irradiation surface should have little effect on the location of the irradiation surface \citep[see][]{Watanabe08}. Nevertheless, the vertical variation of the temperature, and more importantly of the thermal relaxation time, could have a more critical impact on the time evolution of the disk temperature. We discuss this point in more detail in section~\ref{sec:evolution}.}

The vertical hydrostatic approximation is valid as long as the disk's thermal relaxation time $t_{\rm th}$ (see equation~\eqref{eq:tth} for  definition) is much longer than the orbital timescale $t_{\rm K} = 2\pi/\Omega_{\rm K}$. In protoplanetary disks, the condition $t_{\rm th} \gg t_{\rm K}$ is typically fulfilled  at $r \ll 100~\rm au$ \citep[][see also section~\ref{sec:Ueda21_results}]{Dullemond00,Watanabe08,Wu21} but can break down farther out. As discussed by \citet{Wu21}, the disk's hydrodynamic response may influence its thermal and dynamical evolution. In this study, we concentrate on the relatively inner disk region where the vertical hydrostatic assumption is applicable.

The opacities used in our model depend on the dust-to-gas mass ratio and size distribution of the opacity-dominating dust grains. One can account for dust settling by using vertically varying opacities.

\subsection{Disk temperature evolution}\label{sec:evolution}
Because passively irradiated disks can be thermally unstable, we treat the disk interior temperature as intrinsically time-dependent.
For simplicity, we neglect radial heat advection by accreting gas \citep[see, e.g.,][]{Cannizzo93}, radial diffusion of the disk's own thermal radiation \citep[e.g.,][]{Latter12,Owen14}, and accretion heating. We plan to include these effects in future work.

Neglecting the above mentioned effects, the time evolution of the interior temperature $T$ obeys the following energy equation, 
\beq
\frac{\gamma+1}{2(\gamma-1)}\frac{\kB \Sigma}{m_g}\frac{\pd T}{\pd t} = 
2{\cal C}[F_{\rm rep, \downarrow} + \sigma_{\rm SB}(T_{\rm ex}^4 - T^4)],
\label{eq:T_evol}
\eeq
where $\gamma$ is the adiabatic index (taken to be $1.4$ throughout this paper), $T_{\rm ex}$ is the radiation temperature of the parent molecular cloud \citep{Ueda21},
and ${\cal C}$ is a dimensionless factor correcting for the effects of the disk's infrared optical thickness and albedo.

{Equation~\eqref{eq:T_evol} is the vertically integrated equation of total energy conservation \citep{Watanabe90,Watanabe08}. Its} left-hand side stands for the rate of change in the total energy per unit disk area. {The temperature on the left-hand side originally stands for} the density-weighted vertical average of the temperature \citep{Watanabe90}.  {Following \citet{Watanabe08}, we have applied the vertical isothermal approximation and represented the vertically averaged temperature with the single interior temperature. In reality, the rate of change in temperature in optically thick disks depends on the depth from the disk surface, with shallower and optically thinner regions having shorter cooling timescales. Recently, \citet{Pavlyuchenkov22} have pointed out that the vertical variation of the thermal relaxation timescales could weaken the thermal wave instability around the midplane of optically thick disks. This effect is not included in our current modeling.} 

The right-hand side of Equation~\eqref{eq:T_evol} accounts for radiative heating and cooling on both sides ($z< 0$ and $z>0$) of the disk. 
We take the correction factor ${\cal C}$ to be 
\beq
{\cal C} = 
\frac{4 \sqrt{\eps}\tanh(\tau_{\rm eff,mid})}{\sqrt{3} +2\sqrt{\eps} \tanh(\tau_{\rm eff,mid})}
\frac{1}{1+3\tau_{\rm R,mid}/4},
\label{eq:C}
\eeq
where 
\beq
\eps = \frac{\kappa_{\rm P}}{\chi_{\rm R}},
\eeq
is the ratio between the Planck-mean absorption opacity $\kappa_{\rm P}$ {(see equation~\eqref{eq:kappaP} for definition)} and Rosseland extinction opacity $\chi_{\rm R}$, 
\beq
\tau_{\rm eff,\rm mid} =\int_0^\infty \sqrt{3\eps}\, \chi_{\rm R}\rho\, dz' =\int_0^\infty \sqrt{3\chi_{\rm R}\kappa_{\rm P}}\,\rho\, dz'
\label{eq:taueffmid}
\eeq
is the effective absorption optical depth to the midplane accounting for multiple scattering \citep{Rybicki79}, and
\beq
\tau_{\rm R,mid} = \int_0^\infty \chi_{\rm R}\rho\, dz'
\label{eq:tauRmid}
\eeq
is the Rosseland-mean vertical optical depth to the midplane. All the mean opacities appearing here are for the disk thermal radiation.
In equation~\eqref{eq:C}, the factor involving $\tau_{\rm eff,mid}$ corrects for the infrared emissivity and absorptivity of both optically thick and thin disks, derived by approximating a disk with a uniform slab (see equation~\eqref{eq:emissivity} in appendix~\ref{appendixC}). 
The factor $(1+3\tau_{\rm R,mid}/4)^{-1}$ corrects for the vertical radiative diffusion flux in optically thick  ($\tau_{\rm R,mid} \gg 1$) disks being inversely proportional to $\tau_{\rm R,mid}$ \citep{Wu21}. 

\subsection{Numerical implementation}\label{sec:numerics}
We consider a radial computational domain spanning $r= r_{\rm min}$ to $r = r_{\rm max}$ and discretize it into logarithmically spaced cells. Each radial cell has the inner and outer boundaries at 
$r_{j-1/2} = r_{\rm min} (r_{\rm max}/r_{\rm min})^{(j-1)/J}$ and $r_{j+1/2} = r_{\rm min} (r_{\rm max}/r_{\rm min})^{j/J}$, respectively, and the logarithmic center at $r_j = \sqrt{r_{j-1/2}r_{j+1/2}} = r_{\rm min} (r_{\rm max}/r_{\rm min})^{(j-1/2)/J}$, where $j = 1,2,\dots J$ labels the cells and $J$ is the number of the cells. 
One should take the cell width should to be sufficiently smaller than the width $\sim h$ of the weighting function $W$.

The irradiation surface is found with a ray-tracing approach. We use rays emanating from the coordinate origin $(r,z) = (0,0)$ at angle $\theta$ with respect to the $r$-axis.  
All simulations presented in this paper  adopt 180 linearly spaced $\theta$ grids spanning $\theta = 0$ to $\theta = \theta_{\rm max} = \pi/12$, with a resolution of $d\theta = \theta_{\rm max}/180 = 1.45\times 10^{-3}$. We have also tried simulations with $d\theta = \theta_{\rm max}/360$ and confirmed that the higher angular resolution gives no appreciable change in the simulation results.

When searching for the irradiation surface, one must assume the starlight optical depth to the inner computational boundary. The starlight optical depth $\tau_*$  (equation~\eqref{eq:taustar}) to arbitrary radial position $r$ along a ray with angle $\theta$ can be written as
\beq
\tau_* = \tau_{*,\rm in} + \frac{1}{\cos\theta}\int_{r_{\rm min}}^r \chi_* \rho(r',r' \tan\theta) dr',
\label{eq:taustar_numerical}
\eeq
where $\tau_{*,\rm in}$ denotes the optical depth to the inner computational boundary. The second term in the right hand side of equation~\eqref{eq:taustar_numerical} uses $z = r \tan\theta$ and $ds = dr /\cos\theta$. 
The problem here is that $\tau_{*,\rm in}$ is intrinsically unknown because it is determined by the disk structure outside the computational domain. However, as demonstrated in appendix~\ref{appendixD}, an unreasonable choice of $\tau_{*,\rm in}$ can cause unwanted artifacts in the resulting temperature distribution near the inner boundary. 
{Our prescription for $\tau_{*,\rm in}$ is presented in appendix~\ref{appendixD}.} 

The energy equation~\eqref{eq:T_evol} is solved with a first-order forward differencing scheme.
The time step must be smaller than the {thermal relaxation timescale for the vertically averaged temperature} \citep{Watanabe08},
\beq
t_{\rm th} = \frac{\gamma+1}{2(\gamma-1)}\frac{\kB \Sigma}{m_g \sigma_{\rm SB}T^3{\cal C}}.
\label{eq:tth}
\eeq
Note that $t_{\rm th}$ depends on the correction factor ${\cal C}$ introduced in section~\ref{sec:evolution}.

Because the radial computational domain is limited to $r_{\rm min} < r' < r_{\rm max}$, the radial integration in equation~\eqref{eq:Fs} cannot be extended to $r' < r_{\rm min}$ and $r' > r_{\rm max}$. However, neglecting the reprocessed starlight from outside the computational domain would result in an underestimate of the  temperature near the computational boundaries. Near the inner computational boundary, an underestimated temperature amplifies the numerical wiggle in the radial disk structure (see appendix~\ref{appendixD}).  To mitigate the underestimation of the temperature, we account for the flux outside the computational domain in an approximate way. Specifically, we write $F_{\rm rep, \downarrow}$ as 
\beq
F_{\rm rep, \downarrow} = F_{\rm rep, \downarrow,\rm domain} + F_{\rm rep, \downarrow, in} + F_{\rm rep, \downarrow, out},
\label{eq:Fs_decomp}
\eeq
where
$F_{\rm rep, \downarrow,\rm domain}$ represents the flux from within the computational domain,
whereas $F_{\rm rep, \downarrow, in}$ and $F_{\rm rep, \downarrow, out}$ represent the additional fluxes from $r'<r_{\rm min}$ and $r'>r_{\rm max}$, respectively. {Our prescription for these additional fluxes is described in appendix \ref{appendixD}.}

To suppress numerical instabilities arising from grid-scale fluctuations of the reprocessed starlight flux, we replace $F_{\rm rep, \downarrow}(r_j)$ at $j = 2, 3, \dots, J-1$ with the geometric means of the raw fluxes at $j-1$ and $j+1$, $\sqrt{F_{\rm rep, \downarrow}(r_{j-1})F_{\rm rep, \downarrow}(r_{j+1})}$. 
At $j = 1$ and $J$, we use $\sqrt{F_{\rm rep, \downarrow}(r_{1})F_{\rm rep, \downarrow}(r_{2})}$ and $\sqrt{F_{\rm rep, \downarrow}(r_{J-1})F_{\rm rep, \downarrow}(r_{J})}$ instead.
We expect that this grid-scale smoothing would become unnecessary once we include the radial diffusion of the disk's thermal emission in equation~\eqref{eq:T_evol} in future work. 

\section{Validation with gapped disks}\label{sec:validation}
In this section, we test our global two-layer radiative transfer model against disks with an annular surface density gap.
Specifically, we examine if the global two-layer model reproduces the temperature profiles of gapped disks previously obtained by \cite{Jang-Condell12} using a Monte Carlo radiative transfer model.

\subsection{Model}
\begin{figure*}[t]
\begin{center}
\includegraphics[width=16.3cm, bb=0 0 641 584]{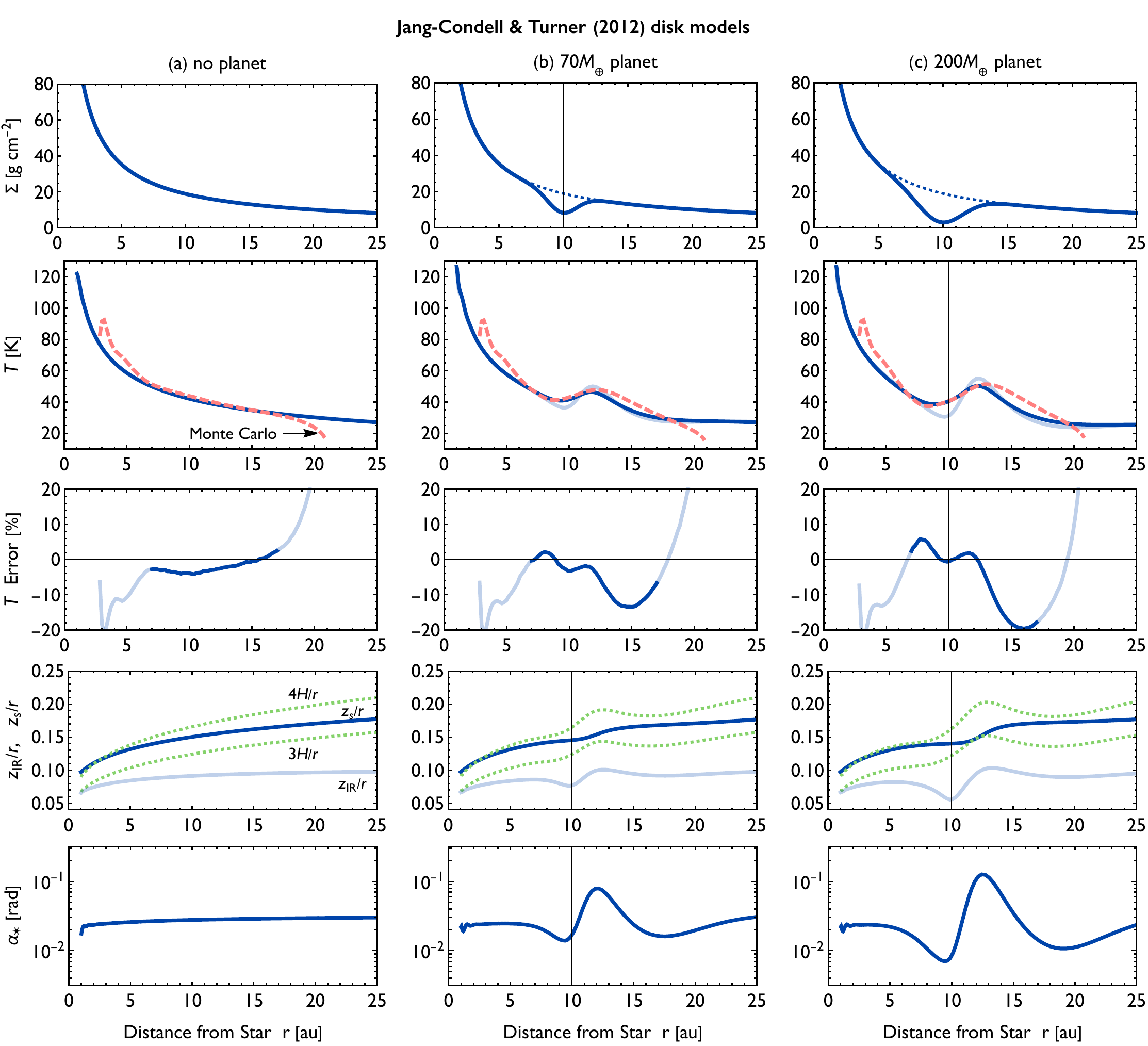}
\end{center}
\caption{Steady-state irradiation structure of the \citet{Jang-Condell12} disk models, reproduced with global two-layer radiative transfer calculations. Columns (a), (b), (c) are for the disk models with no planet, a 70 $M_\oplus$ planet at 10 au, and a 200 $M_\oplus$. The planet is at 10 au. Top row: assumed gas surface density profiles (solid lines) compared with the no-planet profile (dashed). Second row: steady-state temperature profiles {(thick solid lines)} and equilibrium temperatures from Monte Carlo radiative transfer calculations \citep[{red} dashed lines]{Jang-Condell12}. The thin solid lines give temperatures from the local two-layer model, equation~\eqref{eq:T_parallel_2}, using $\mu_*$ from the global two-layer model. Third row: fractional temperature difference between the global two-layer and Monte Carlo calculations. {Thick and thin lines} show the errors {inside and outside} the region $r \approx 7$--17 au, where the Monte Carlo calculation is least affected by the computational domain boundaries. Fourth row: {the heights of the starlight irradiation surface and infrared thermal photosphere, $z_s$ and $z_{\rm IR}$ (thick and thin solid lines, respectively),} normalized by $r$. {The upper and lower green dotted lines indicate that $z_s \approx 3$--$4H$}.
Bottom row: starlight grazing angle $\alpha_* = \sin^{-1}\mu_*$. 
}
\label{fig:JCT12}
\end{figure*}

Following \citet{Jang-Condell12}, we consider axisymmetric gapped disks with the radial gas surface density profile given by
\beq
\Sigma = 19\pfrac{r}{10~\rm au}^{-0.9}\left[1-d\exp\left(-\frac{(r-a)^2}{2w^2}\right)\right]~\rm g~cm^{-2},
\eeq 
where $19({r}/{10~\rm au})^{-0.9}~\rm g~cm^{-2}$ is the surface density profile with no planet, $a = 10~\rm au$ is the planet's orbital radius, and $d$ and $w$ represent the depth and width of the planet-carving gap, respectively. They considered one case with no planet and two cases with a planet of mass $M_p=70$ or $200M_\oplus$ orbiting at 10 au from the central star. The parameters $(d,w)$ were taken to be $ (0.56, 0.11a)$ and $(0.84, 0.17a)$ for $M_p =$ 70 and 200$M_\oplus$, respectively. 
The top row of figure~\ref{fig:JCT12} shows the gas surface density profiles for three disk models. 
The central star was assumed to have mass $M_* = M_\odot$, radius $R_* = 2.6R_\odot$, surface temperature $T_* = 4280~\rm K$.

\citet{Jang-Condell12} provided the temperature structure of the three disk models ($M_p = 0$, 70, and $200M_\oplus$) using two radiative transfer models. One is a model conceptually similar to ours, relying on the locally one-dimensional analytic solution for a plane-parallel disk \citep{Jang-Condell03,Jang-Condell04,Jang-Condell08,Jang-Condell09}.
The other is the Monte Carlo model developed by \citet{Turner12}, which directly follows the emission, absorption, and scattering of photon packets using the frequency-dependent opacities provided by \citet{Jang-Condell09}.  
The two approaches returned similar results as shown in figure 4 of \citet{Jang-Condell12}, so we only use their Monte Carlo results in the following comparison. 
The dashed lines in the second row of figure~\ref{fig:JCT12} show
the temperature profiles from the Monte Carlo calculations by \citet{Jang-Condell12}.
We note that their Monte Carlo calculations only cover 3--20.8 au.

We adopt the mean opacities for stellar and disk thermal radiation computed from the frequency-dependent opacities of \citet{Jang-Condell09}. For disk thermal radiation, we evaluate the mean opacities at a fixed temperature of $T = 50~\rm K$. The adopted mean opcities are $\chi_* = 11.3~\rm cm^{2}~g^{-1}$, $\kappa_* = 1.36~\rm cm^{2}~g^{-1}$, $\chi_{\rm P} = 1.92~\rm cm^{2}~g^{-1}$, $\chi_{\rm R} = 1.98~\rm cm^{2}~g^{-1}$, and , $\kappa_{\rm P} = 0.975~\rm cm^{2}~g^{-1}$, which yield $\eps_* = 0.12$, $q = 5.7$, $f_\downarrow = {0.24}$ {(see figure~\ref{fig:f_down})}, and $\eps = 0.49$. The disk model considered here is optically thick to its own thermal emission, with $\tau_{\rm R} \sim 10$ at $r \sim 10~\rm au$. The thermal relaxation time at $r\sim 10~\rm au$ is $t_{\rm th} \sim \kB \Sigma/(m_g\sigma_{\rm SB}T^3 \tau_{\rm R}) \sim 10^2~\rm yr$.

Unlike the radiative equilibrium Monte Carlo calculations of \citet{Jang-Condell12}, our global two-layer calculations are fully time-dependent. 
The time step in our calculations is fixed to be 1 yr, which is shorter than the thermal relaxation time in the computational domain. The initial temperature profile is set to be $T= 120(r/1~\rm au)^{-3/7}~\rm K$. This choice is arbitrary, but is not far from the steady state temperature profile for the model with no planet.  

Our two-layer calculations use a computational domain covering $r_{\rm min} = 1$ au to $r_{\rm max} =$ 100 au, divided into 200 logarithmically spaced radial cells. Our computational domain is wider than in the Monte Carlo calculations by \citet{Jang-Condell12}, and therefore our results are likely less affected by the computational boundaries {and by any outer edge to the disk}.
We neglect radiation from the parent molecular cloud by setting $T_{\rm ex} = 0$ in  equation~\eqref{eq:T_evol}. The additional reprocessed starlight fluxes from outside the computational domain, $F_{\rm rep,\downarrow,in}$ and $F_{\rm rep,\downarrow,out}$, are included.

\subsection{Results}\label{sec:JCT12_results}
\begin{figure*}[t]
\begin{center}
\includegraphics[width=16.4cm, bb=0 0 710 322]{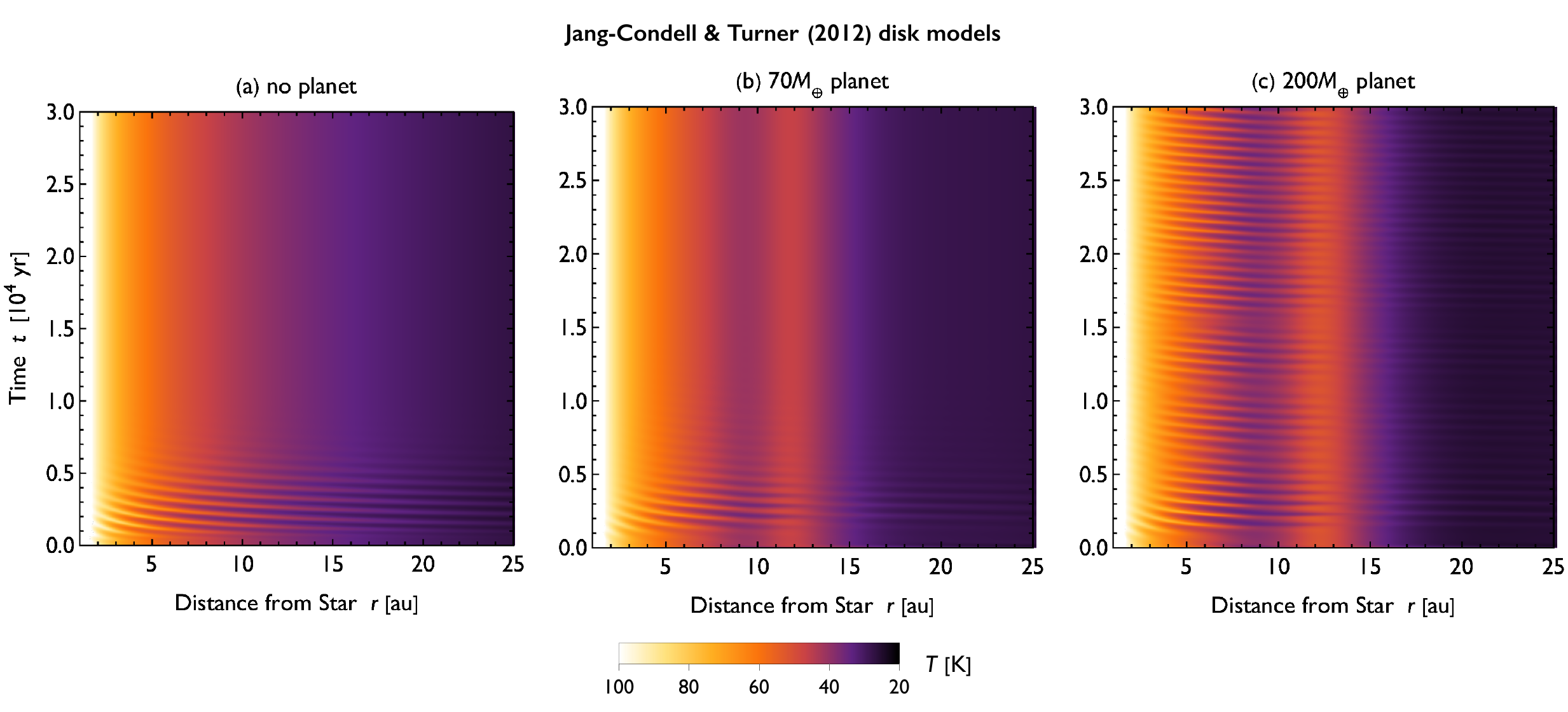}
\end{center}
\caption{Evolution of the temperature structure in time-dependent radiative 
transfer calculations for three \citet{Jang-Condell12} disk models {from figure~\ref{fig:JCT12}}. 
{The temperature oscillates with a period similar to the disk's thermal relaxation timescale, in a pattern that propatates toward the star, likely due to teh thermal wave instability (section~\ref{sec:JCT12_results}).}
}
\label{fig:JCT12_Tt}
\end{figure*}
Figure~\ref{fig:JCT12_Tt} shows the time evolution of the temperature $T$ for three \citet{Jang-Condell12} disk models obtained from our global two-layer calculations. 
In the $M_p = 0$ and $70M_\oplus$ models, the temperature profile relaxes into a steady state. During relaxation, the temperature exhibits a damped oscillation with a period of $\approx 700~\rm yr$. This oscillation period is comparable to the thermal relaxation time of the disk, which is $\sim 200$--400 yr for the model with no planet.  This implies that the observed oscillation is not an artifact but reflects the system's thermal relaxation. In the $M_p =200M_\oplus$ disk model, an oscillation of a similar period persists at a constant amplitude. {The oscillation propagates toward the star, indicating that it is likely due to the thermal wave instability (see section~\ref{sec:twi} for more details about the long-term behavior of this instability)}.
In the following, we describe the results for each disk model in more detail.

\subsubsection{$M_p = 0$ model}
\begin{figure}[t]
\begin{center}
\includegraphics[width=8cm, bb=0 0 288 211]{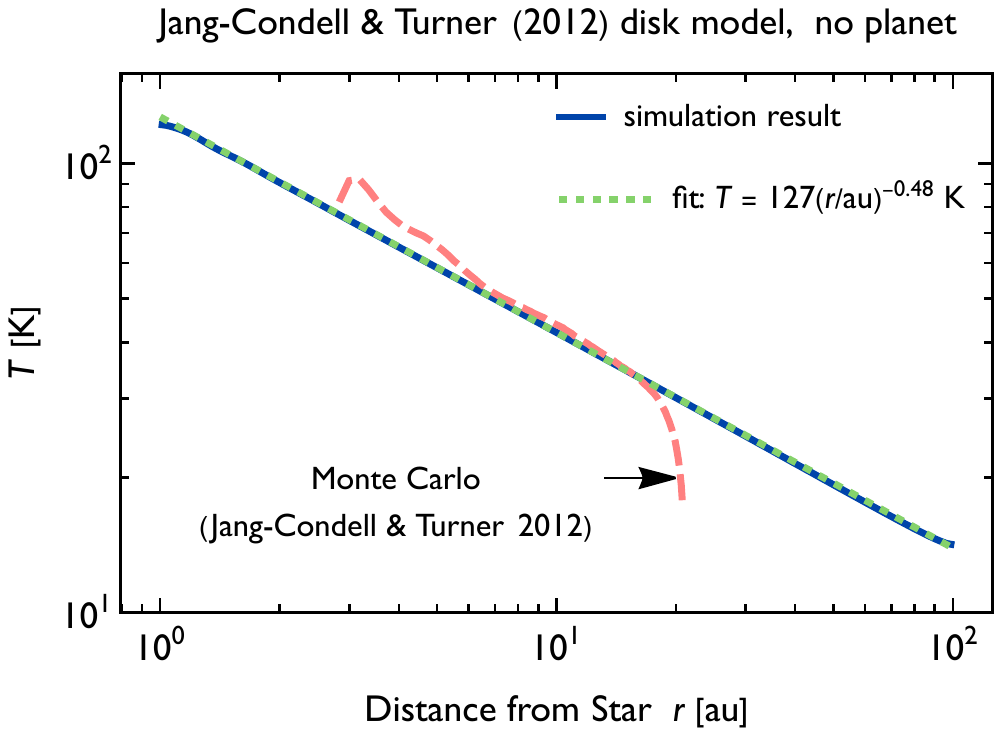}
\end{center}
\caption{Steady-state temperature profile from a global two-layer radiative transfer calculation for the \citet{Jang-Condell12} disk model with no planet {(solid line)} and a power-law fit $T \propto r^{-0.48}$ (dotted line). The dashed line is the equilibrium temperature profile from the Monte Carlo calculation by \citet{Jang-Condell12}.
}
\label{fig:JCT12_model1}
\end{figure}

The steady-state structure of the $M_p = 0$ model at $t = 3\times 10^4$ yr is shown in the left column of figure \ref{fig:JCT12}.
The steady-state temperature profile follows a power law (see figure~\ref{fig:JCT12_model1})\footnote{Our best-fit $T \propto r^{-0.48}$ (see figure~\ref{fig:JCT12_model1}) is  steeper than the well-known profile $T \propto r^{-3/7}$ for optically thick disks with radially constant $z_s/H$ \citep{Chiang97}. In our temperature calculations, the ratio $z_s/H$ generally varies with $r$ because $z_s$ is computed from the radial optical depth. In this no-planet disk model, $z_s$ decreases from $\sim 4H$ at $r \approx 1~\rm au$ to $\sim 3H$ at $r \approx 25~\rm au$ (see the bottom left panel of figure~\ref{fig:JCT12}) and therefore the irradiation surface flares more slowly with $r$ than in constant $z_s/H$ models, explaining the steeper temperature profile.}. This is reasonable because we assume a power-law surface density profile. At $r \approx 7$--$17~{\rm au}$, the steady-state temperature matches the radiative equilibrium temperature from the Monte Carlo calculation by \citet{Jang-Condell12} to within $4\%$ (see the third row of figures~\ref{fig:JCT12}). 
The excellent match can only be achieved when we account for the scattering of stellar radiation;  using $f_\downarrow = 0.5$ (for $\eps_* = 1${; see equation~\eqref{eq:f_down}}) instead of the correct value of $f_\downarrow = {0.24}$ {would} overestimate the temperature by $\approx 40\%$. 
At $r \la 7$ au and $r \ga 17~{\rm au}$, the Monte Carlo result deviates from the power-law profile. This is likely because of the relatively narrow computational domain {and disk outer edge} adopted in the Monte Carlo calculation. In the following, comparisons between our results and those of \citet{Jang-Condell12} are only made at $r \approx 7$--$17~{\rm au}$. 

The irradiation surface lies at $z_s \sim 3$--$4H$ (the fourth row of figures~\ref{fig:JCT12}), and the starlight grazing angle $\alpha_*$ is $\approx 0.02$--0.03 over the entire disk (the bottom row of figure~\ref{fig:JCT12}). The small drop in $\alpha_*$ at the inner computational boundary is an artifact and depends on how we treat  $\tau_{*,\rm in}$ and $F_{\rm rep,\downarrow,in}$. As discussed in appendix~\ref{appendixD}, our prescriptions for $\tau_{*,\rm in}$ and $F_{\rm rep,\downarrow,in}$ already suppress this inner boundary artifact significantly, if not completely. A similar inner boundary artifact can also be seen in the results for $M_p = 70$ and $200M_\oplus$ (see columns (b) and (c) of figure~\ref{fig:JCT12}).

\subsubsection{$M_p = 70M_\oplus$ model}
The steady-state structure of the $M_p = 70M_\oplus$ disk model at $t = 3\times 10^4$ yr is shown in the center column of figure~\ref{fig:JCT12}. The temperature profile has a dip and a bump around $r \approx 9$ and 12 au, approximately corresponding to the bottom and outer edge of the planetary gap, respectively. As already pointed out by \citet{Jang-Condell12}, these features arise because the gap's trough is less exposed to stellar radiation and the gap's outer rim is more illuminated. This can also be seen in the bottom row of figures~\ref{fig:JCT12}, which shows that $\alpha_*$ has a local minimum and a maximum in the gap's trough and outer rim, respectively.

At $r \approx 7$--$17~{\rm au}$, our temperature profile for the $M_p = 70M_\oplus$ model matches the Monte Carlo result by \citet{Jang-Condell12} to an accuracy of $\la 15\%$. {In particular, the error in the gap region} falls below a few \%. The error is larger beyond the gap's outer edge, reaching $15\%$ at $r \sim 15~\rm au$.

\subsubsection{$M_p = 200M_\oplus$ model}\label{sec:200ME}
Because the disk structure for the $M_p = 200M_\oplus$ model is time-dependent due to the thermal wave instability, we generate steady-state disk structure by averaging the simulation result over $t = 1.5$--$3\times 10^4$ yr. The obtained steady-state disk structure  is  shown in the right column of figure~\ref{fig:JCT12}. 
In the inner half of the gap, the starlight grazing angle falls below $10^{-2}$, indicating that this part almost falls into a shadow cast by the inner disk. 
As in the case of $M_p = 70M_\oplus$, our two-layer model reproduces the gap temperature from the Monte Carlo calculation by \citet{Jang-Condell12} to an accuracy of a few \%. The agreement is less good beyond the gap's outer edge, where the error approaches $20\%$. 

\begin{figure}
\begin{center}
\includegraphics[width=8cm, bb=0 0 288 325]{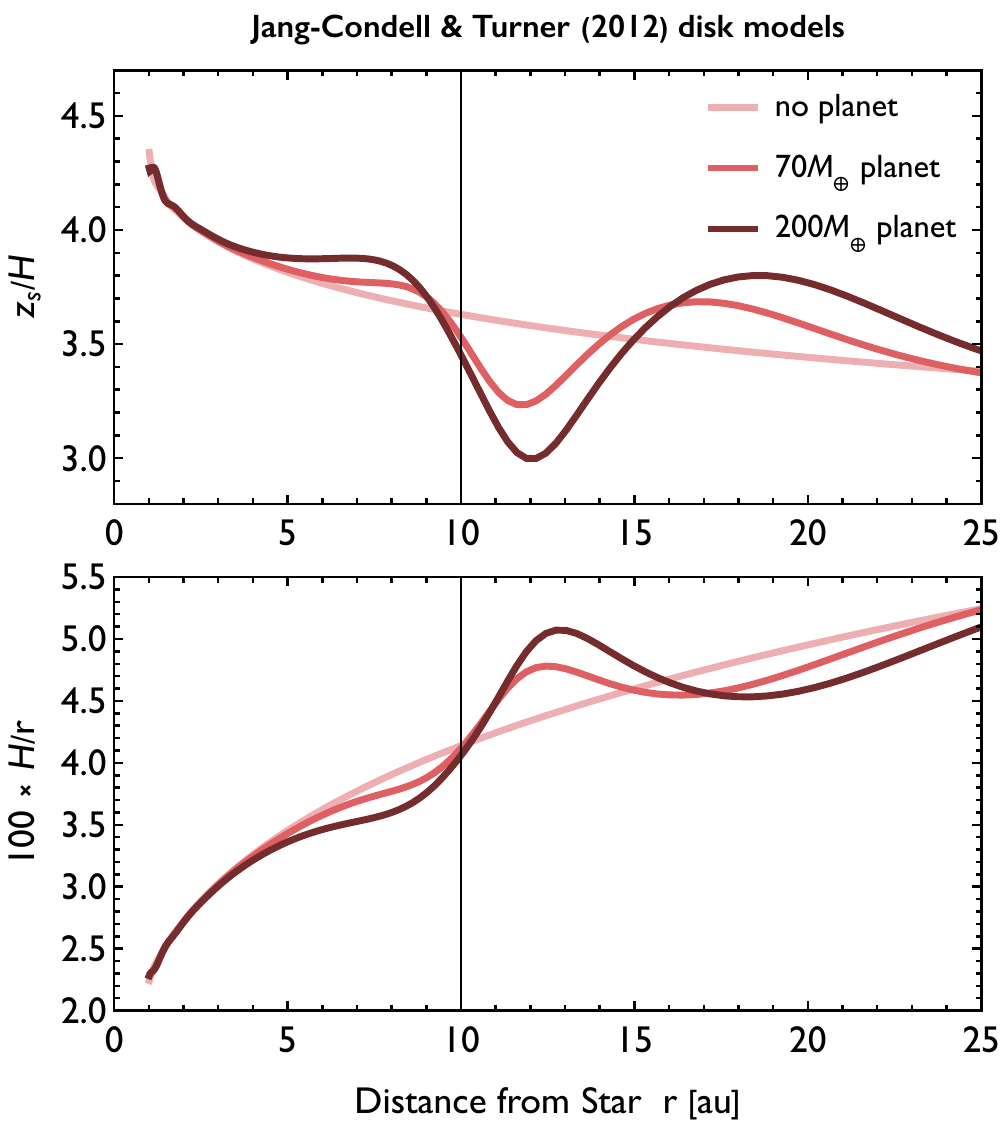}
\end{center}
\caption{Steady-state radial profiles of $z_s/H$ and $H/r$ (upper and lower panels) obtained using the global two-layer method for the three \citet{Jang-Condell12} disk models {from figure~\ref{fig:JCT12}}. Note that high values of $z_s/H$ are associated with plateaus of $H/r$ \citep{Garaud07,Wu21}.
}
\label{fig:JCT12_zsH}
\end{figure}
It is interesting to ask why the $M_p = {200}M_\oplus$ model is thermally unstable while the other two models are not. In general, the thermal wave instability operates in disks with large $z_s/H$ \citep{Wu21,Ueda21}. As shown in the upper panel of figure~\ref{fig:JCT12_zsH}, the $M_p = {200}M_\oplus$ model is indeed the one among the three models that has the largest $z_s/H$, except around the gap's sunlit outer edge ($r \sim 12~\rm au$). 
In this model, the deep gap modifies the disk structure such that $H/r$ becomes radially flat at $r \sim 8$ au and $r \sim 20$ au (see the lower panel of figure~\ref{fig:JCT12_zsH}). As shown by \citet{Garaud07} and \citet{Wu21}, a radially flat $H/r$ generally leads to a high value of $z_s/H$. We suspect that the relatively high $z_s/H$ around the ${200}M_\oplus$ planet's gap destabilizes the disk. 

\subsubsection{The role of radial radiation transfer}
We have shown that our global two-layer model reproduces the temperature profiles from the Monte Carlo calculations by \citet{Jang-Condell12} to within an accuracy of 20\%. Before closing this section, we also show that local two-layer models would never produce such an accurate temperature profile for gapped disks. In the second row of figure~\ref{fig:JCT12}, we overplot the temperature profiles one would have from the local two-layer model, equation~\eqref{eq:T_parallel_2}, using the profiles of $\mu_*$ derived from our global two-layer calculations. It can be seen that the local model appreciably underestimates and overestimates the temperatures at the gap's trough and outer rim, respectively. This suggests that our global treatment of the reprocessed starlight is essential for accurately predicting the gap temperature.

\section{Application to the thermal wave instability}\label{sec:twi}
The thermal wave instability is one of the most interesting targets of our global two-layer  radiative transfer model. The previous simulations by \citet{Watanabe08}, \citet{Ueda21}, and \citet{Wu21} showed that the nonlinear stage of the instability is characterized by a train of inward moving temperature peaks with extended shadows.
However, the simulations by \citet{Watanabe08} and \citet{Ueda21} used a simplified radiative transfer model with ad hoc radial smoothing for reprocessed starlight, which may not accurately resolve the thermal waves. Moreover, their simulations overestimated the disk cooling rate in optically thick regions as pointed out by \citet{Wu21}. In contrast, \citet{Wu21} used a Monte Carlo radiative transfer code that fully includes the radial radiation transfer of reprocessed starlight. However, they only simulated the relatively early phase of the instability over several relaxation times.

In this section, we use our global two-layer model to simulate the long-term ($t \gg t_{\rm th}$) evolution of the thermal wave instability. The aims here are to study how the delayed thermal relaxation in optically thick regions, radial radiative transfer of reprocessed starlight, and radial numerical resolution affect the nonlinear development of the instability. We describe the model in section~\ref{sec:Ueda21_model} and present the results in section~\ref{sec:Ueda21_results}. 

\subsection{Model}\label{sec:Ueda21_model}
We adopt the disk model used by \citet{Ueda21}. The gas surface density profile is given by $\Sigma = 1700(r/1~{\rm au})^{-3/2}\exp(-r/100~\rm au)~g~cm^{-2}$. The disk opacities are assumed to scale with the dust-to-gas mass ratio $f_{\rm d2g}$ of opacity-dominating dust grains. For the fiducial value of $f_{\rm d2g} = 0.01$, the mean opacities are $\chi_{*} = \kappa_{*} = 8~\rm cm^2~g^{-1}$ and $\chi_{\rm R} =\chi_{\rm P} = \kappa_{\rm P} = 4~\rm cm^2~g^{-1}$. for dust grains is a free parameter of the model. In this study, we only consider this fiducial model. Because $\eps_* = 1$, we have $f_\downarrow = 0.5$. External thermal radiation of $T_{\rm ex} = 10~\rm K$ is included, so that the disk temperature never falls below 10 K.

The radial computational domain ranges between 0.03 to 300 au and is divided into 480 logarithmically spaced cells, resulting in a radial resolution of $dr/r = 0.019$. The adopted radial resolution is two times better than in the simulation by \citet{Ueda21}. In section~\ref{sec:Ueda21_resolution}, we also present simulation runs with lower resolution to study the convergence of our simulation results. 

Our simulations differ from those by \citet{Ueda21} in the treatment of radial radiative transfer and thermal relaxation. To discriminate between the effects of the two different treatments, we also conduct a simulation using the same local radiative transfer model as in the simulations by \citet{Ueda21}, but including the correction for thermal relaxation in optically thick regions, the factor $(1+3\tau_{\rm R,mid}/4)^{-1}$ in equation~\eqref{eq:C}.
Their radiative transfer model, originally developed by \citet{Watanabe08}, approximates the downward reprocessed starlight flux as
\beq
F_{\rm rep, \downarrow}(r) = \frac{L_*}{8\pi}\left\langle \frac{A_s(r')}{r'^2} + \frac{4R_*}{3\pi r'^3}\right\rangle,
\label{eq:Fs_W08} 
\eeq
\beq
A_s (r) = 1-\exp\left(-\int_{z_s(r)}^\infty \kappa_{*}\rho (r,z')dz'\right),
\label{eq:As}
\eeq
where the angled brackets denote a radial average over a radial zone near $r'=r$.
This averaging is introduced to mimic the radial propagation of reprocessed starlight over distance $\sim z_s(r)$.
Following \citet{Ueda21}, we use the Gaussian weight function  $\exp[-(r'-r)^2/z_s(r)^2]$ for the radial averaging in equation~\eqref{eq:Fs_W08}.

The simulations are carried out over 1 Myr. The simulated time covers $\sim 100 t_{\rm th}$ at 10 au and $\sim 10t_{\rm th}$ at 1 au (see section~\ref{sec:Ueda21_results}).
In comparison, the simulation reported by \citet{Wu21} only covers several thermal times at 10 au and less than one thermal time at 1 au.
Therefore, our simulation for the first time fully captures thermal waves traveling from the 10 au to 1 au regions, with a correct treatment for thermal relaxation in the optically thick regime. 
The timestep $\Delta t$ is taken to be 1 yr, which is 10 times shorter than the minimum thermal relaxation time in the disk {($\sim 10~\rm yr$; see figure~\ref{fig:Ueda21_tth} in section~\ref{sec:Ueda21_results})}.
Using a shorter timestep of $\Delta t = 0.25$ yr makes no appreciable change in the simulation results.

All our simulations assume  vertical hydrostatic equilibrium (see section~\ref{sec:vertical}). As shown in section~\ref{sec:Ueda21_results}, the requirement $t_{\rm th} \gg  t_{\rm K}$ for the hydrostatic equilibrium is fulfilled over the entire thermally unstable region.

\subsection{Results}
\label{sec:Ueda21_results}
\begin{figure*}
\begin{center}
\includegraphics[width=16.4cm, bb=0 0 576 477]{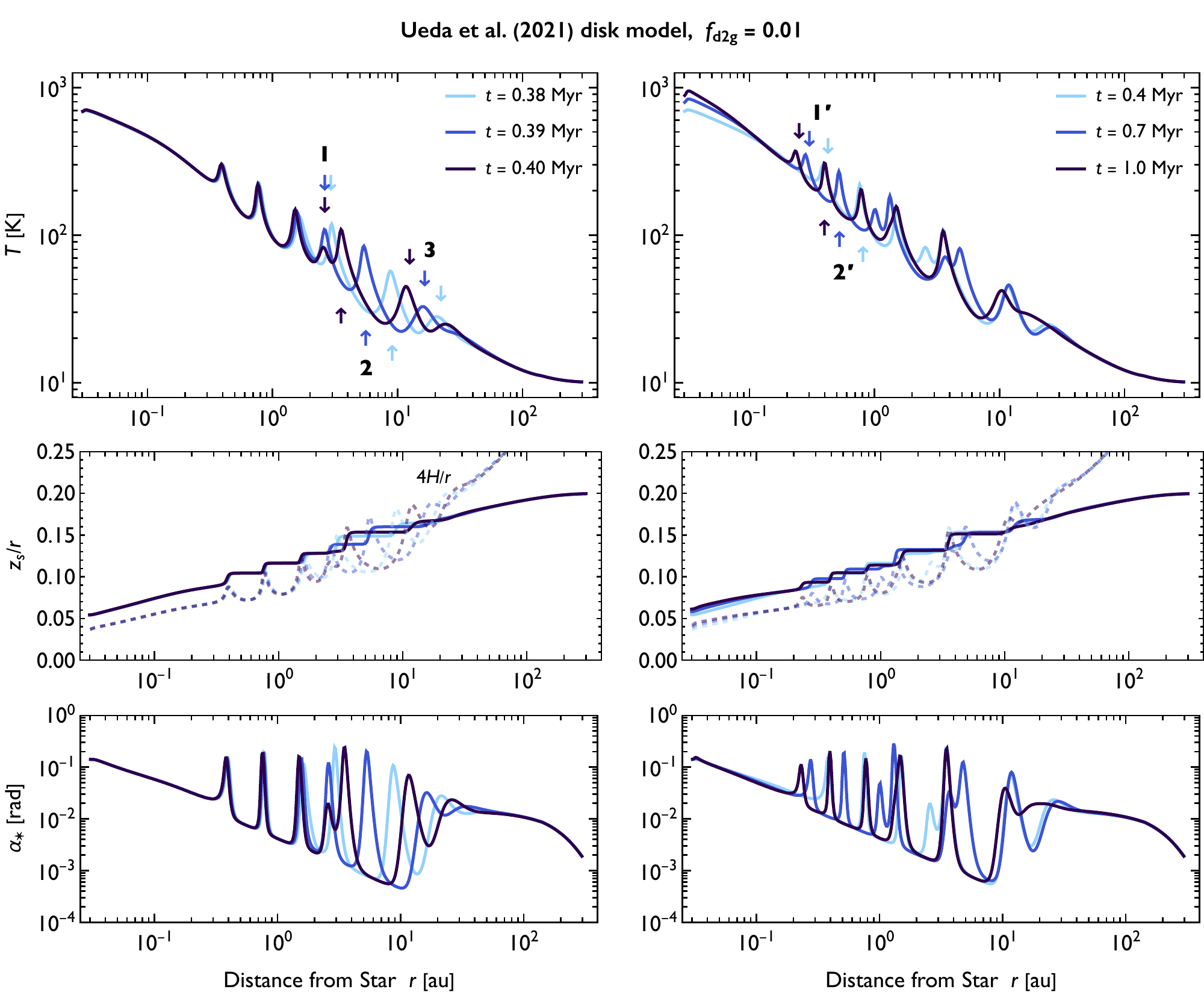}
\end{center}
\caption{Radial profiles of the disk interior temperature $T$ (top panels), normalized irradiation surface height $z_s/r$ (middle panels), and starlight grazing angle $\alpha_*$ (bottom panels) from the global two-layer calculation for the \citet{Ueda21} disk model with $f_{\rm d2g} = 0.01$. The left and right columns show the time variation in the intervals $t=0.38$--0.40 Myr and 0.4--1.0 Myr, respectively. The arrows in the top panels mark five selected temperature peaks (1, 2, 3, 1$'$, and 2$'$) migrating inward. At $t = 0.40$ Myr,  peak 2 catches up with peak 1 and starts merging with it.  The dotted lines in the middle panels indicate $4H/r$.
}
\label{fig:Ueda21}
\end{figure*}
Figure~\ref{fig:Ueda21} presents snapshots of the temperature distribution at selected times. 
We find that the temperature structure at $r \approx 0.3$--30 au is unstable and exhibits oscillations (thermal waves) that propagate inward. 
The thermal waves consist of  sharp temperature peaks with width $\sim z_s(r_{\rm peak})$, where $r_{\rm peak}$ is the radial position of each peak. 
Each peak casts a shadow with a radial width $\sim r_{\rm peak}$ (see the middle and bottom panels of figure~\ref{fig:Ueda21}), and each shadow causes a dip in the temperature profile. 
The irradiation surface height in the unstable region exceeds $4H$, which is higher than in the \citet{Jang-Condell12} disk models. This provides further support for the prediction by \citep{Wu21} that disks with larger $z_s/H$ are more prone to the  thermal wave instability.

\begin{figure}
\begin{center}
\includegraphics[width=8cm, bb=0 0 288 211]{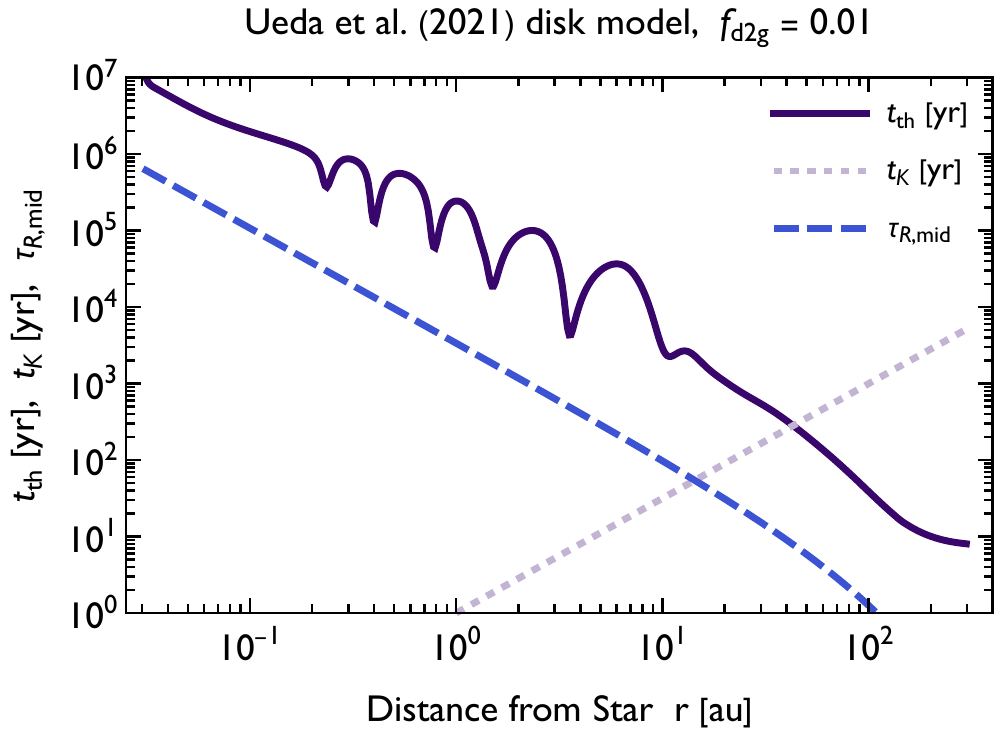}
\end{center}
\caption{Radial profiles of the thermal relaxation timescale $t_{\rm th}$ (equation~\eqref{eq:tth}; solid line), {local Keplerian time (dotted line)}, and vertical infrared optical depth to the midplane, $\tau_{\rm R, mid} =\kappa_{\rm R}\Sigma/2$ ({dashed} line), for the \citet{Ueda21} disk model with $f_{\rm d2g} = 0.01$. The thermal relaxation time shown here uses the temperature distribution at $t = 1~\rm Myr$ obtained from our simulation (see the top right panel of figure~\ref{fig:Ueda21}). 
}
\label{fig:Ueda21_tth}
\end{figure}

At $r \ga 30~\rm au$ and $r \la 0.3~\rm au$, the thermal wave instability is suppressed for different reasons. In the outer region of $r \ga 30~\rm au$, the external radiation flux $\sigma_{\rm SB}T_{\rm ex}^4$ is comparable to or even dominates over the reprocessed starlight flux $F_{\rm rep, \downarrow}$, directly suppressing the thermal wave instability  \citep{Ueda21}. In the inner region of $r \la 0.1~\rm au$, the finite size of the central star determines the starlight grazing angle (i.e., $\alpha_* \approx 4R_*/(3\pi r)$; see equation~\eqref{eq:mustar}), which stabilizes the thermal wave instability as it is triggered by the variation of the stellar grazing angle with temperature fluctuations. In addition, the thermal timescale in this inner region is too long for thermal waves to develop within 1 Myr. Indeed, figure~\ref{fig:Ueda21_tth}, shows that the thermal relaxation time $t_{\rm th}$ (equation~\eqref{eq:tth}) at $r \la 0.1~\rm au$ exceeds 1 Myr. 
Figure~\ref{fig:Ueda21_tth} also shows that the condition $t_{\rm th} \gg t_{\rm K}$ for vertical hydrostatic equilibrium is fulfilled in the entire region where the thermal wave instability is observed.

\begin{figure}[t]
\begin{center}
\includegraphics[width=8cm, bb=0 0 267 401]{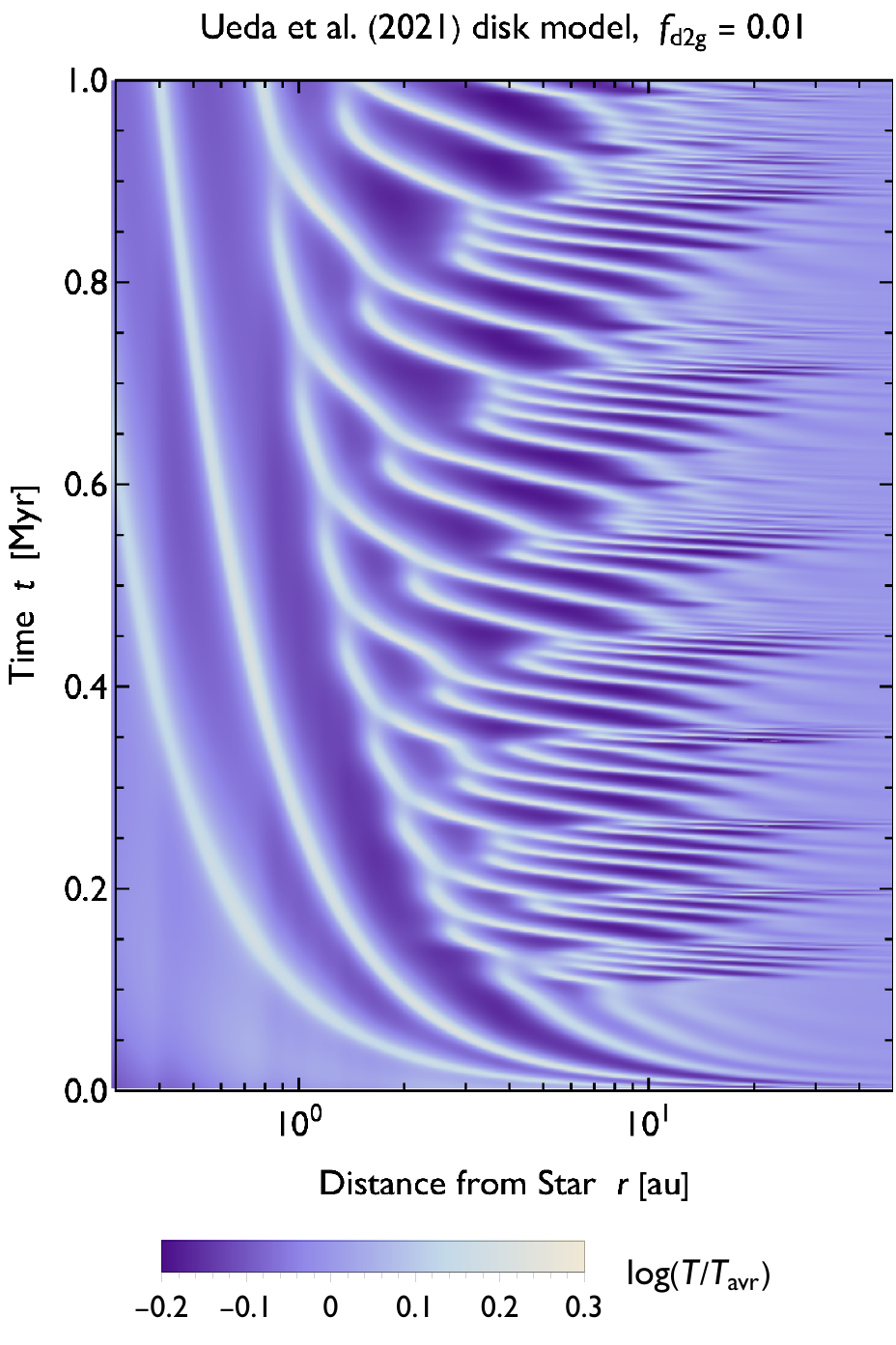}
\end{center}
\caption{Spacetime plot for the temperature $T$, normalized by the time average $T_{\rm avr}(r)$ over $t = 0.1$--1 Myr, from the global two-layer calculation for the \citet{Ueda21} disk model with $f_{\rm d2g} = 0.01$. This plot highlights the thermal waves.
}
\label{fig:Ueda21_wave}
\end{figure}
Compared to the thermal waves observed in the simulations by \citet[][see their figures 2 and 3]{Ueda21}, our thermal waves are less coherent and propagate inward on different timescales at different radial positions.
This can be more clearly seen in figure~\ref{fig:Ueda21_wave}, which shows a spacetime plot of the temperature distribution.
To highlight the wave components, we here normalize the temperature profile by the time-average $T_{\rm avr}(r)$ over $t = 0.1$--1 Myr.  
One can see that temperature peaks at smaller $r$ migrate inward on longer timescales. The observed migration timescale at each $r$ is crudely consistent with the local thermal timescale $t_{\rm th}$ shown in figure~\ref{fig:Ueda21_tth}. This suggests that the radial variation of the migration speed is a manifestation of the radial variation of $t_{\rm th}$. In contrast, in the simulations by \citet{Ueda21}, $t_{\rm th}$ was nearly independent of $r$ because they did not include the correction for the thermal relaxation rate. This explains why the thermal waves observed by \citet{Ueda21} propagate inward on a radially uniform timescale. For the \citet{Ueda21} disk model with $f_{\rm d2g} = 0.01$, the thermal relaxation correction must be included because the disk is optically thick ($\tau_{\rm R,\rm mid} > 1$) at $r \la 100~\rm au$ (see figure~\ref{fig:Ueda21_tth}).

Another interesting feature that was not visible in the simulations by \citet{Ueda21} is the collision of temperature peaks. As an example, the top left panel of figure~\ref{fig:Ueda21} shows how two temperature peaks labelled by 1 and 2 collide. The outer peak 2 migrates faster than the inner peak 1, and they merge together at $t \approx 0.40~\rm Myr$.
We speculate that the radially varying migration timescales of the temperature peaks induce their collisions.

Because the observed temperature peaks are narrow, it is important to examine whether the finite numerical resolution affects our simulation results. {In appendix~\ref{sec:Ueda21_resolution}, we show that a radial resolution of $dr/r \la 0.04$ is enough to resolve the thermal waves.} 

\begin{figure}[t]
\begin{center}
\includegraphics[width=8cm, bb=0 0 267 417]{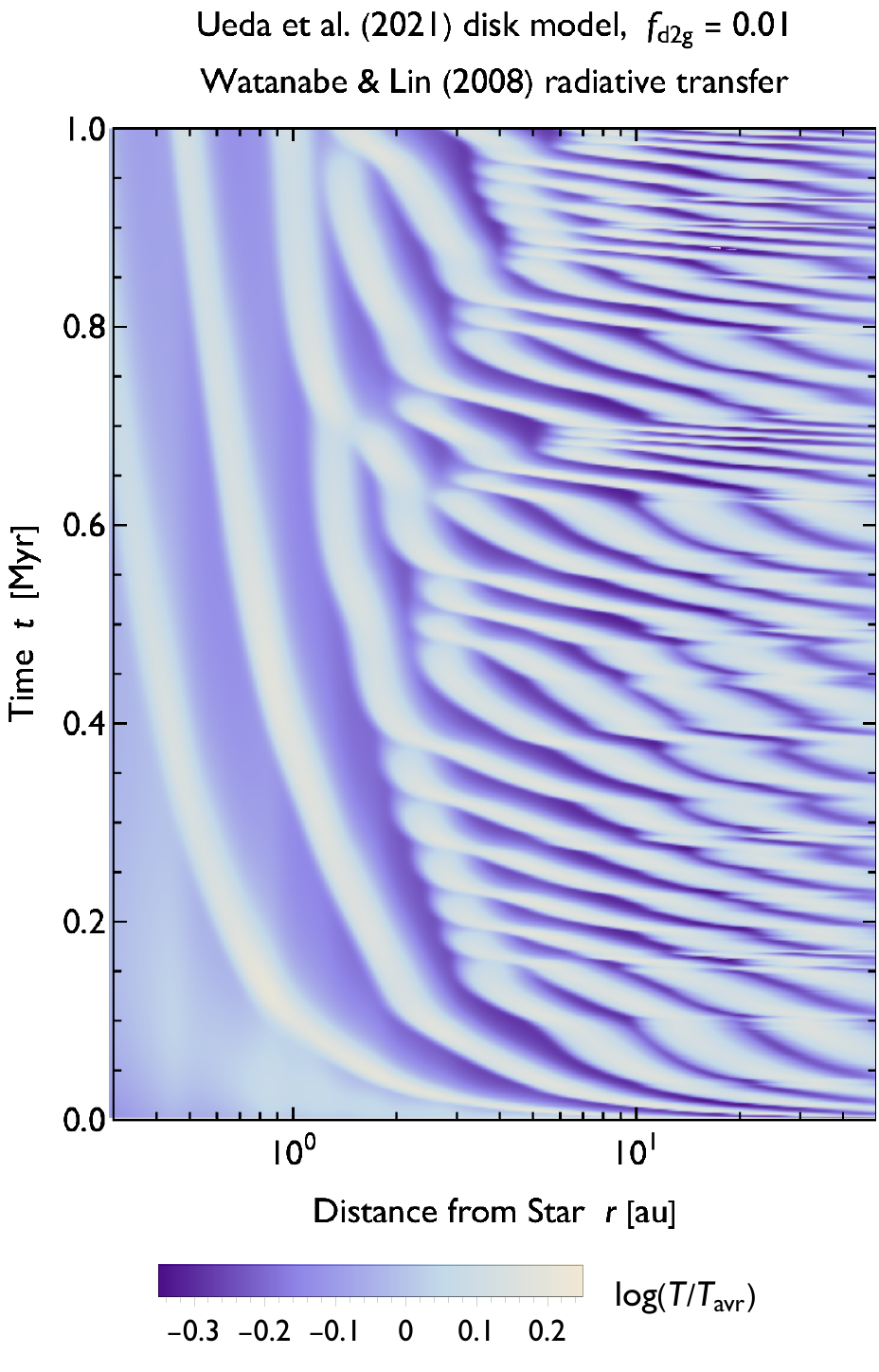}
\end{center}
\caption{
Same as figure~\ref{fig:Ueda21_wave}, but from a simulation using the  \citet{Watanabe08} radiative transfer model with Gaussian radial smoothing of the reprocessed starlight. 
}
\label{fig:Ueda21_W08}
\end{figure}
Figure~\ref{fig:Ueda21_W08} shows the spacetime temperature plot from the simulation using the radiative transfer model of \citet{Watanabe08}.
The results are qualitatively similar to those from the global two-layer simulation presented above in that both feature collisions of temperature peaks.
However, the \citet{Watanabe08} model produces wider  temperature peaks than our global two-layer model. 
This is likely due to the ad hoc radial averaging of the reprocessed starlight flux adopted in the model.
Therefore, we conclude that the radiative transfer model of \citet{Watanabe08} captures the qualitative nature of the thermal wave instability but should not be used for a quantitative study of the instability.

\subsection{Implications for dust evolution and planetesimal formation}\label{sec:implications}
\begin{figure}
\begin{center}
\includegraphics[width=8cm, bb=0 0 267 357]{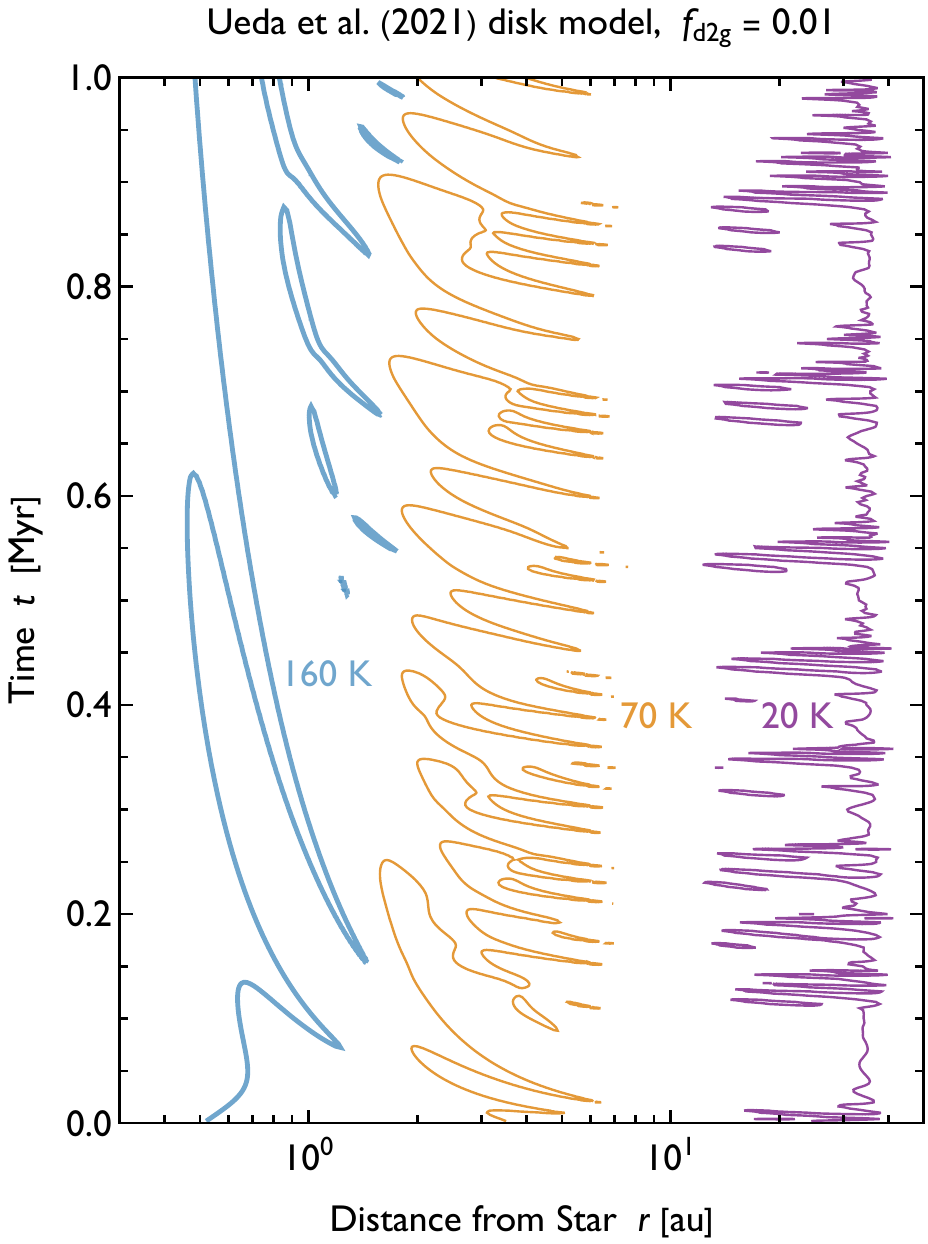}
\end{center}
\caption{Spacetime plot showing how the positions where $T =$ 160 K (thick lines), 70 K (center thin lines), and 20 K (right thin lines) vary with time, obtained from the simulation for the \citet{Ueda21} disk model with $f_{\rm d2g} = 0.01$. These positions correspond to the snow lines of H$_2$O, CO$_2$, and CO ices, respectively. }
\label{fig:Ueda21_snowlines}
\end{figure}

As already noted by previous studies \citep{Watanabe08,Ueda21,Wu21}, the thermal wave instability may have two important implications for dust evolution and planetesimal formation in dust-rich disks. First, the thermal wave instability causes temporal variations in snow line locations.  We demonstrate this in figure~\ref{fig:Ueda21_snowlines}, where we show how the positions where $T = $ 160, 70, and 20 K move with time in the simulation shown in figures~\ref{fig:Ueda21} and \ref{fig:Ueda21_wave}. 
The selected temperatures correspond to the sublimation temperatures of H$_2$O, CO$_2$, and CO ices, respectively \citep{Okuzumi16}.
Overall, the plot indicates that the instability causes order-of-unity time variations in the positions of the snow lines. For instance, the H$_2$O snow line would migrate between $\approx 0.4$--1.5 au on the timescale of $\sim 0.1~\rm Myr$.
The previous simulation by \citet{Ueda21} predicted a migration timescale of $\sim 10~\rm yr$ for the  H$_2$O snow line. Our results indicate that the simulations by \citet{Ueda21} underestimated the migration timescale of the  H$_2$O snow line because their simulations did not include the correction for thermal relaxation in the optically thick limit.
Our new simulation suggests that the H$_2$O snow line can in fact move on a timescale comparable to the planet formation timescale, $\sim 0.1$--1 Myr, if the disk's  optical thickness is large enough.
We expect that the snow line migration should have important effects on planetesimal formation around the snow line \citep{Ros13,Schoonenberg17,Drazkowska17,Ida21,Hyodo21}, in particular on the composition of forming planetesimals. 
In future work, we will explore the potential effects by including dust evolution in our global two-layer radiative transfer calculations.  

\begin{figure}
\begin{center}
\includegraphics[width=8cm, bb=0 0 288 208]{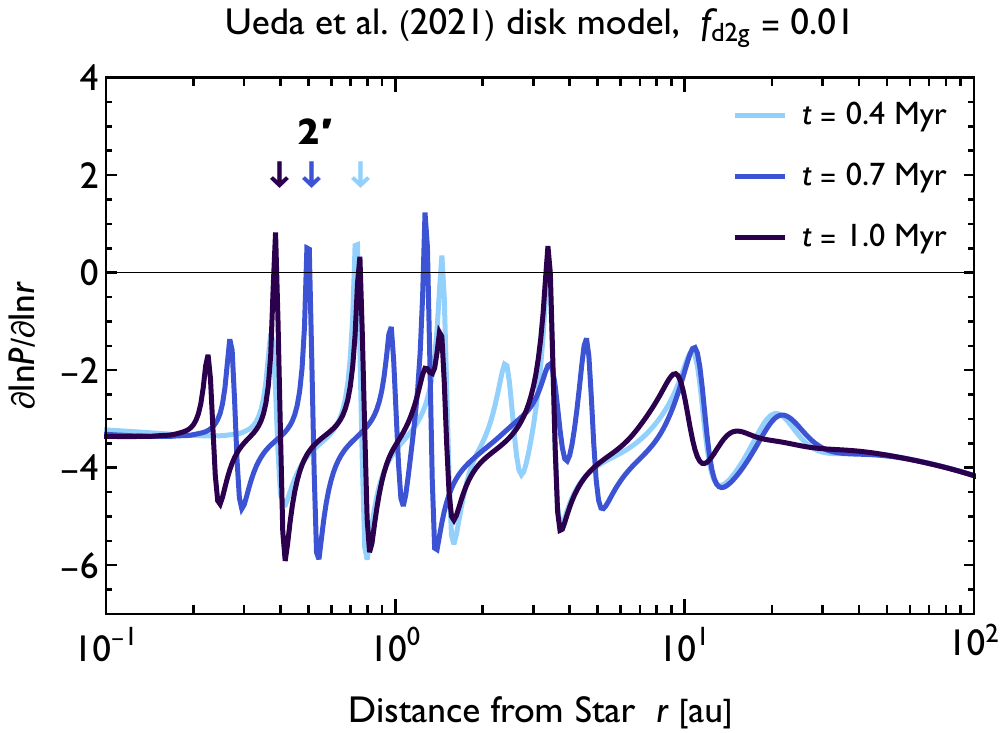}
\end{center}
\caption{
Radial pressure gradient $\pd\ln P/\pd\ln r$ at the midplane as a function of $r$ at different times, obtained from the global two-layer simulation presented in figure~\ref{fig:Ueda21}.
The arrows  indicate the positions of temperature peak 2$'$ shown in the top right panel of figure~\ref{fig:Ueda21}.}
\label{fig:Ueda21_dlnPdlnr}
\end{figure}

Second, the thermal wave instability might produce pressure maxima that can trap dust particles.
Dust particles in a gas disk are known to drift in the direction of increasing gas pressure \citep{Whipple72,Adachi76,Weidenschilling77a}.  
Specifically, the radial drift velocity of a dust particle in a gas disk is given by 
\beq
v_{d,r} = \frac{\Omega_{\rm K} t_{\rm stop}}{1+(\Omega_{\rm K} t_{\rm stop})^2}\frac{c_s^2}{v_{\rm K}}\frac{\pd \ln P}{\pd \ln r},
\label{eq:vdr}
\eeq
where $t_{\rm stop}$ is the stopping time of the particle, $v_{\rm K} = r \Omega_{\rm K}$ is the local Keplerian velocity, and $P = \rho \kB T/m_g$ is the gas pressure.
The radial drift velocity is proportional to the radial pressure gradient 
${\pd\ln P}/{\pd\ln r} = {\pd\ln \rho}/{\pd\ln r} +  {\pd\ln T}/{\pd\ln r}$, 
with negative and positive radial gradients leading to inward and outward drift, respectively.
A pressure maximum acts as a trap for radially drifting particles because their drift velocity converges there. 
\citet{Watanabe08} already suggested that the thermal wave instability can potentially produce local pressure maxima. Our new simulations relying on a more rigorous radiative transfer model confirm this possibility. 
Figure~\ref{fig:Ueda21_dlnPdlnr} shows some snapshots of the radial distribution of $\pd\ln P/\pd\ln r$ at the midplane from our global two-layer simulation presented in Figure~\ref{fig:Ueda21}. 
We find that some of the temperature peaks observed in figure~\ref{fig:Ueda21} (e.g., peak 2$'$ shown in the right panel of figure~\ref{fig:Ueda21}) indeed reverse the sign of the pressure gradient locally.
Less pronounced temperature peaks that do not reverse the pressure gradient may also slow down particle inward drift and thereby promote planetesimal formation. 
However, a question remains as to whether the steep pressure variations observed here are hydrodynamically stable, because they may violate the Rayleigh criterion for stable disk rotation \citep[e.g.,][]{Yang10}.  {More fundamentally, the actual thermal waves around the midplane of optically thick disks could be less pronounced than predicted from our calculations owing to the vertically varying thermal relaxation time \citep{Pavlyuchenkov22}.}
Addressing {these issues} is beyond the scope of this paper but should be done in future work. 

\section{Summary}\label{sec:summary}
We have presented a new two-layer radiative transfer model for computing the radial temperature distribution of axisymmetric, irradiated protoplanetary disks. 
Unlike the standard two-layer model, our new model explicitly accounts for the radial transfer of reprocessed starlight and is therefore applicable to disks with shadowed regions. 
Our global two-layer model is conceptually similar to the radiative transfer models of \citet{Jang-Condell03,Jang-Condell04} and \citet{Jang-Condell08}, but is much simpler and computationally more efficient thanks to the assumed disk axisymmetry.
In addition, our model treats the radial temperature distribution as time-dependent (section~\ref{sec:evolution}), allowing us to study disks' thermal instabilities.  

We have tested the global two-layer model against the Monte Carlo radiative transfer calculations by \citet{Jang-Condell12} for gapped disks (section~\ref{sec:validation}).
We have found that the steady-state temperature profiles from our global two-layer calculations match the previous Monte Carlo results to within an accuracy of 20\% even with a deep gap carved by a 200$M_\oplus$ planet (figure~\ref{fig:JCT12}). We have also found that disks with larger $z_s/H$ are more prone to the thermal wave instability, confirming the prediction from the linear analysis by \citet{Wu21}.

We have then used the global two-layer model to simulate, for the first time, the long-term evolution of the thermal wave instability  with a correct treatment for thermal relaxation in the optically thick limit (section~\ref{sec:twi}).
Our simulations reveal that the thermal waves are highly chaotic, with inward migrating temperature peaks colliding and merging frequently (figures~\ref{fig:Ueda21} and \ref{fig:Ueda21_wave}).
Our simulations also show that the migration timescale of the temperature peaks increases as they move inward, reaching $\ga 0.1~\rm Myr$ at 1 au. These observed properties of the thermal waves are likely attributed to the radially varying thermal relaxation timescale in our simulations.
Snow lines also migrate on a timescale similar to the local thermal relaxation time (figure~\ref{fig:Ueda21_snowlines}).
Sharp temperature peaks produced by the thermal wave instability can reverse the sign of the radial pressure gradient locally (figure~\ref{fig:Ueda21_dlnPdlnr}), indicating that they may act as a dust trap. 
Our future modeling will include dust evolution and study the coupled evolution of dust and disk temperature structure in detail.  

Finally, we note that our current treatment of the disk vertical density structure is greatly simplified, relying on the isothermal and hydrostatic approximations (section~\ref{sec:vertical}). The isothermal approximation may introduce some errors in the evaluation of the irradiation surface height, {although the errors appear to be small \citep{Watanabe08}}. The hydrostatic approximation is inapplicable to the outer disk region where the thermal relaxation time is shorter than the orbital time. Adopting a two-temperature vertical density profile \citep{Watanabe08} and treating the disk scale height as a dynamical variable \citep{Dullemond00} may allow us to relax these assumptions without having to fully solve the hydrodynamic equation of motion. We will pursue this direction in future work. 
Including accretion heating, radial diffusion of the disk's own thermal radiation, {and vertically varying thermal relaxation timescales} will be other important future directions. 

\begin{ack}
The authors thank Tilman Birnstiel, Cornelis Dullemond, Mario Flock for helpful discussions and Ralf Siebenmorgen for a careful review of the manuscript. This work was supported by JSPS KAKENHI Grant Numbers JP18H05438, JP19J01929, JP19K03926, JP20H00182, and JP20H01948. This work was carried out in part at the Jet Propulsion Laboratory, California Institute of Technology, under contract with NASA and with the support of NASA grant 18-2XRP18\_2-0059.
\end{ack}

\bibliography{irradiation}

\appendix

\section{Fraction of starlight reprocessed downward}\label{appendixA}
Here, we derive the fraction of the stellar radiation reprocessed downward, $f_\downarrow$ (equation~\eqref{eq:f_down}), 
using the locally plane-parallel disk model of \citet{Calvet91}.

We consider stellar (visible) radiation incident on a plane-parallel atmosphere at angle $\alpha_* = \sin^{-1}\mu_*$ with respect to the atmosphere's surface.
Particles in the atmosphere either absorb or scatter the incident light and reprocess the absorbed component into infrared radiation. 
Part of the multiple-scattered starlight escapes from the atmosphere and the rest is eventually converted into infrared radiation.
\citet{Calvet91} derived the mean intensities and vertical fluxes of the multiple-scattered starlight and disk thermal emission in a locally plane-parallel disk in radiative equilibrium. They used the first and second moments of the frequency-integrated radiative transfer equations for the visual and infrared radiation with the Eddington approximation.
\citet{Jang-Condell04} extended the model of \citet{Calvet91} so that the model can deal with infrared scattering. 
The moment equations for the scattered starlight read \citep[see also equations~(10) and (11) of][]{DAlessio98}
\beq
\frac{dJ_s}{d\tau_s} = \frac{3}{4\pi} F_s,
\label{eq:dJsdtaus}
\eeq
\beq
\frac{dF_s}{d\tau_s} = 4\pi \eps_* J_s - (1-\eps_*) F_*e^{-\tau_s/\mu_*},
\label{eq:dFsdtaus}
\eeq
where $J_s$ and $F_s$ are the frequency-integrated mean intensity and vertical upward flux for the scattered starlight, respectively, $\tau_s = \int_z^\infty \chi_* \rho dz'$ is the vertical extinction optical depth, and $\eps_*$ is the ratio between the absorption and extinction opacities $\kappa_*$ and $\chi_*$ for the stellar radiation (see equation~\eqref{eq:epsstar}). {We follow \citet{DAlessio98} and evaluate $\kappa_*$ and $\chi_*$ using the Planck means at the stellar surface temperature $T_*$, 
\beq
\kappa_* = \frac{\pi}{\sigma_{\rm SB} T_*^4} \int_0^\infty \kappa_\nu B_\nu(T_*) d\nu,
\label{eq:kappa_star}
\eeq
\beq
\chi_* = \frac{\pi}{\sigma_{\rm SB} T_*^4} \int_0^\infty \chi_\nu B_\nu(T_*) d\nu,
\label{eq:chi_star}
\eeq
where $\kappa_\nu$ and $\chi_\nu$ are the monochromatic absorption and extinction opacities at frequency $\nu$, respectivey, and $B_\nu$ is the Planck function.  }

The moment equations for the disk's infrared radiation are \citep[see also equations~(15) and (16) of][]{DAlessio98}
\beq
\frac{dJ_d}{d\tau} = \frac{3}{4\pi}F_d,
\label{eq:dJddtau}
\eeq
\beq
\frac{dF_d}{d\tau} = 4\pi \eps (J_d - B_d),
\label{eq:dFddtau}
\eeq
where $J_d$, $F_d$, and $B_d$ are the frequency-integrated mean intensity, vertical upward flux, and Planck function for the disk thermal radiation, respectively, $\tau = \int_z^\infty \chi \rho dz'$ is the mean vertical optical depth for the infrared radiation, and $\eps = \kappa/\chi$.
The mean opacities $\kappa$ and $\chi$ appering here are formally defined as the intensity- and flux-weighted averages of the absorption and extinction opacities for the disk thermal radiation, respectively. Following \citet{DAlessio98}, we approximate $\kappa$ and $\chi$ with the Planck and Rosseland mean opacities, 
\beq
\kappa_{\rm P} = \frac{\pi}{\sigma_{\rm SB} T^4} \int_0^\infty \kappa_\nu B_\nu(T) d\nu,
\label{eq:kappaP}
\eeq
\beq
\chi_{\rm R} = \frac{4\sigma_{\rm SB} T^3}{\pi \int_0^\infty \chi_\nu^{-1} {(dB_\nu(T)}/{dT}) d\nu}
\label{eq:chiR}
\eeq
 (see \citealt{Hubeny03} for the validity of these choices at large optical depths).

To fix the boundary conditions, we assume that both $F_s$ and $F_d$ vanish at large optical depths and that the outgoing radiation at the disk surface is hemispherically isotropic,
\beq
F_s(\tau_s=0) = 2\pi J_s(\tau_s=0),
\label{eq:bc_s0}
\eeq
\beq
F_d(\tau=0) = 2\pi J_d(\tau=0).
\label{eq:bc_d0}
\eeq

The resulting $J_d(\tau_s)$ is given by equation (12) of \citet{Calvet91}. 
For $\mu_* \ll 1$ and $\tau_{\rm eff} \equiv \sqrt{3\epsilon} \tau \gg 1$, the expression for $J_d$ can be greatly simplified as
\beq
J_d \approx \frac{1}{2\pi}
\left[\eps_* + 
\left(2\sqrt{3\eps_*}+\frac{3}{q}\right)\frac{1-\eps_*}{3+2\sqrt{3\eps_*}} \right] 
\mu_* F_*.
\eeq
Because the thermal radiation at these depths is nearly isotropic,
we may decompose it into upward and downward components that are hemispherically isotropic.
The downward component has a flux of magnitude
$F_{\rm rep,\downarrow} = \pi J_{d}$, which can be rewritten as $F_{\rm rep,\downarrow} = f_\downarrow \mu_* F_*$ if we define $f_\downarrow$ by equation~\eqref{eq:f_down}.

\section{Reprocessed starlight flux for plane-parallel disks}\label{appendixB}
In this section, we prove that our expression for $F_{\rm rep, \downarrow}$ (equation~\eqref{eq:Fs_2}) recovers the more familiar expression $F_{\rm rep, \downarrow} = f_\downarrow \mu_* F_*$ (equation~\eqref{eq:Fs_planeparallel}) in the limit where
$h$ is small compared to the other relevant radial length scales. 
In  this limit, the Lorentzian factor $1/( (r'-r)^2 +h^2)$ in the weighting function $W$ is sharply peaked at $r' = r$, and therefore the other factors inside the radial integration in equation~\eqref{eq:Fs_2} can be approximately evaluated at $r' = r$. We thus obtain   
\beqn
F_{\rm rep, \downarrow}(r) 
\approx  
 \frac{ f_\downarrow(r) \mu_*(r)F_*(r) h_{r=r'} } {\pi } \int_0^\infty \frac{dr'}{(r'-r)^2 +h_{r=r'}^2} ,~~
 \label{eq:Fs_B}
 \eeqn
 where we have used that $h \ll r \approx r'$, $E(1) \approx 1$, 
 and $\sqrt{1+(dz_s/dr')^2} \approx 1$. 
The radial integration in equation~\eqref{eq:Fs_B} yields 
 \beq
 \int_0^\infty \frac{dr'}{(r'-r)^2 +h_{r=r'}^2} 
 = \frac{1}{2h_{r=r'} }\left(\pi + 2\tan^{-1}\pfrac{r}{h_{r' =r}}\right),
 \label{eq:int}
 \eeq
For $h_{r' =r} \ll r$, we have $\tan^{-1}(r/h_{r' =r}) \approx \pi/2$, which yields $F_{\rm rep, \downarrow} \approx f_\downarrow \mu_* F_*$ .

\section{Numerical treatment of the computational boundaries}\label{appendixD}
As note in section~\ref{sec:model}, our radiative transfer calculations require the starlight optical depth to the inner computational boundary, $\tau_{*,\rm in}$, and the reprocessed starlight fluxes from outside the inner and outer computational boundaries, $F_{\rm rep,\downarrow,in}$ and $F_{\rm rep,\downarrow,out}$.   
All these depend on the conditions outside the computational domain and therefore must be assumed. 

For $\tau_{*,\rm in}$, we adopt the following prescription
\beq
\tau_{*,\rm in} = A_\tau \chi_* \rho(r_{\rm min},r_{\rm min} \tan\theta) \frac{r_{\rm min}}{\cos\theta}.
\label{eq:taustarin}
\eeq
Here, ${r_{\rm min}}/{\cos\theta}$ stands for the distance from the coordinate origin to the position $(r,r \tan\theta)$. The dimensionless number $A_\tau$ is a free parameter and is expected to be of order unity for uniformly-flared disks with smooth radial density structure. 
\citet{Flock16} adopted a similar prescription for $\tau_{\rm *,in}$ with $A_\tau = 1$.
However, as shown below, the choice $A_\tau = 1$ introduces a small wiggle in the temperature distribution near the inner computational boundary for the simulations presented in section~\ref{sec:validation}. 
Throughout this paper, we adopt $A_\tau = 0.3$, which better suppresses the wiggle for these particular simulations.

For $F_{\rm rep,\downarrow,in}$ and $F_{\rm rep,\downarrow,out}$, we approximate the irradiation surface at $r'<r_{\rm min}$ and $r>r_{\rm max}$ with flat planes lying at heights $z_s(r_{\rm min})$ and $z_s(r_{\rm max})$, extending radially over $0<r'<r_{\rm min}$ and $r'>r_{\rm max}$, and producing vertical fluxes of the reprocessed starlight per irradiation surface area of $[f_\downarrow \mu_* F_*]_{r'=r_{\rm min}}$ and $[f_\downarrow \mu_* F_*]_{r'=r_{\rm max}}$ respectively.
With these simplifications, the additional fluxes can be analytically calculated as (see equations~(20) and (21) of \citealt{Ueda17} for a derivation) 
\beqn
F_{\rm rep, \downarrow, in}(r) 
&=& \frac{[f_\downarrow \mu_* F_*]_{r'=r_{\rm min}}}{\pi} \int_0^{r_{\rm min}} 
\frac{h_{r'=r_{\rm min}}dr'}{h_{r'=r_{\rm min}}^2+(r-r')^2}
\nonumber \\
&=& \frac{[f_\downarrow \mu_* F_*]_{r'=r_{\rm min}}}{\pi} 
\left[\tan^{-1}\pfrac{r}{h_{r' = r_{\rm min}}} 
-\tan^{-1}\pfrac{r-r_{\rm min}}{h_{r' = r_{\rm min}}}\right],
\nonumber \\
\label{eq:Fs_in}
\eeqn
\beqn
F_{\rm rep, \downarrow, out}(r) 
&=& \frac{[f_\downarrow \mu_* F_*]_{r'=r_{\rm max}}}{\pi} \int_{r_{\rm max}}^\infty 
\frac{h_{r'=r_{\rm max}}dr'}{h_{r'=r_{\rm max}}^2+(r-r')^2}
\nonumber \\
&=& \frac{[f_\downarrow \mu_* F_*]_{r'=r_{\rm max}}}{\pi} 
\left[\frac{\pi}{2}-\tan^{-1}\pfrac{r_{\rm max}-r}{h_{r' = r_{\rm max}}}\right].
\label{eq:Fs_out}
\eeqn

\begin{figure*}
\begin{center}
\includegraphics[width=8cm, bb=0 0 288 305]{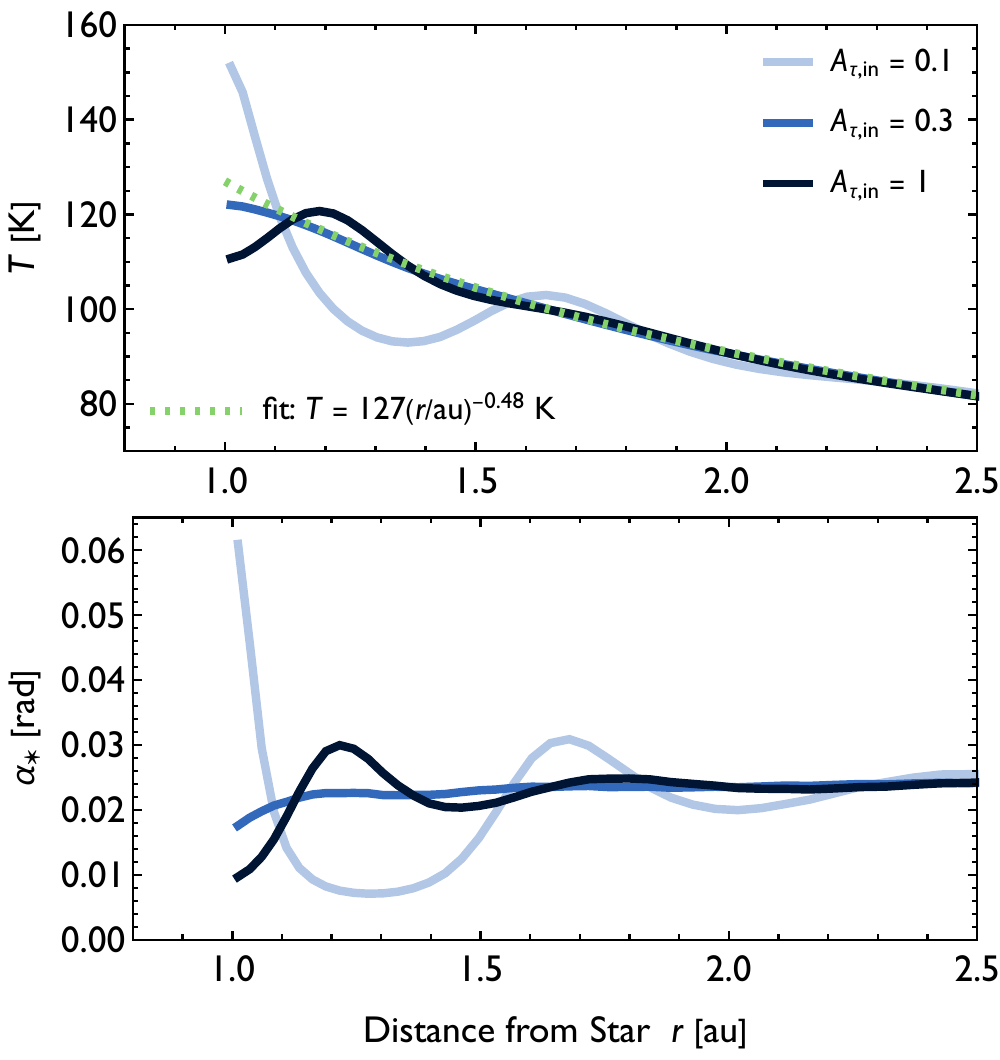}
\hspace{2mm}
\includegraphics[width=8cm, bb=0 0 288 305]{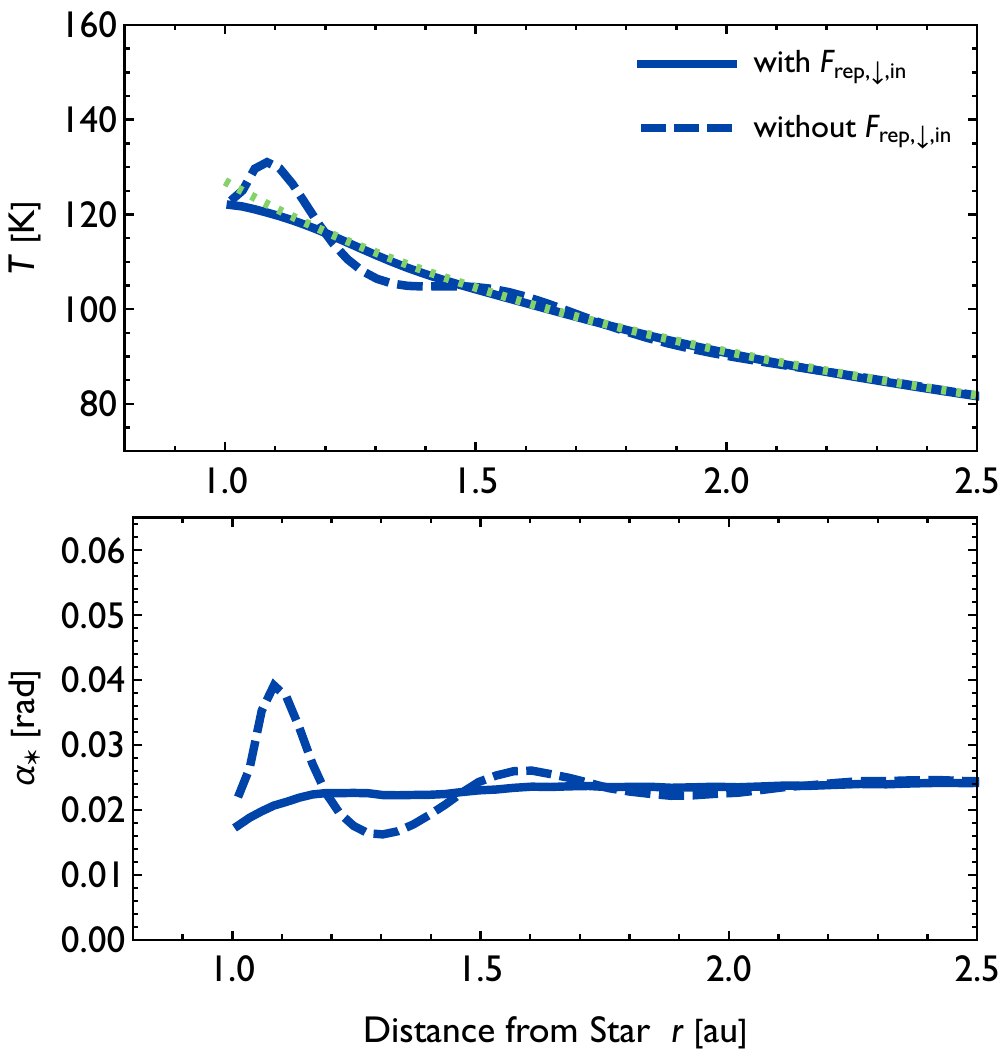}
\end{center}
\caption{Steady-state disk structure of the \citet{Jang-Condell12} disk model with no planet obtained from global two-layer calculations with different prescriptions for the inner boundary. The upper and lower panels zoom in on the interior temperature $T$ and starlight grazing angle $\alpha_{*}$, respectively, near the inner computational boundary. The left column compares the calculations with three different values of the parameter $A_{\tau, \rm in}$ controlling the starlight optical depth to the inner boundary (see equation~\eqref{eq:taustarin}), with $A_{\tau, \rm in} = 0.3$ being the default value. 
The right column compares the calculations for $A_{\tau, \rm in} = 0.3$ with and without the additional reprocessed flux beyond the inner boundary, $F_{\rm rep,\downarrow,in}$ (equation~\eqref{eq:Fs_in}).
The dotted line in the upper panels is a power-law fit for the bulk temperature profile in the computational domain shown in figure~\ref{fig:JCT12_model1}. 
}
\label{fig:inner}
\end{figure*}
\begin{figure}
\begin{center}
\includegraphics[width=8cm, bb=0 0 288 167]{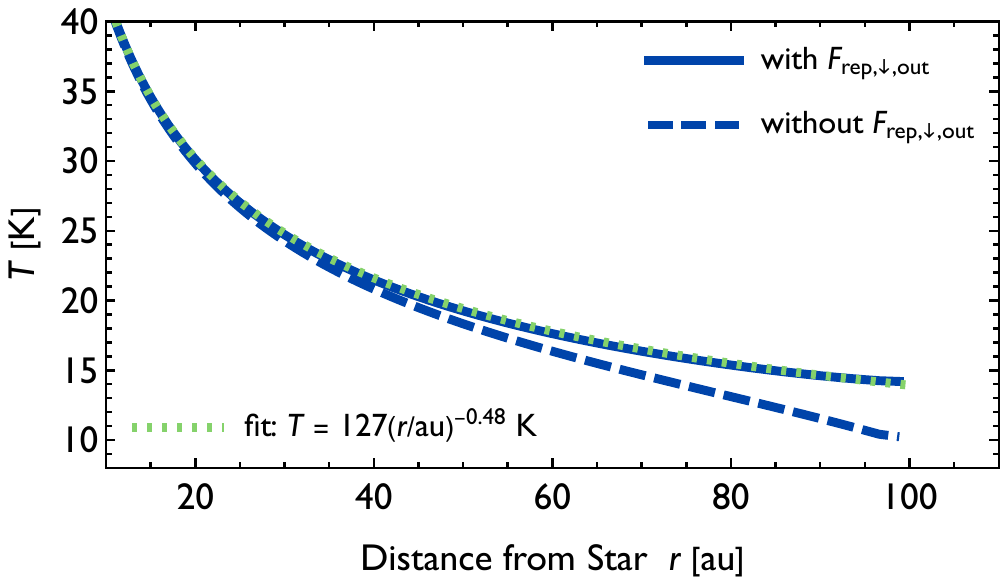}
\end{center}
\caption{Steady-state temperature distribution in the outer region of the \citet{Jang-Condell12} disk with no planet from global two-layer calculations with and without the additional reprocessed flux beyond the out boundary, $F_{\rm rep,\downarrow,out}$ (equation~\eqref{eq:Fs_out}).
The dotted line in the temperature plot is a power-law fit for the bulk temperature profile in the computational domain shown in figure~\ref{fig:JCT12_model1}. 
}
\label{fig:out}
\end{figure}
Below, we examine how our prescriptions for $\tau_{*,\rm in}$, $F_{\rm rep,\downarrow,in}$, and $F_{\rm rep,\downarrow,out}$   
affect the calculated disk structure near the boundaries. 
We select the \citet{Jang-Condell12} disk model with no planet for a case study. The left column of figure~\ref{fig:inner} shows the steady-state structure near the inner boundary of this disk model calculated with different values of $A_{\tau,\rm in}$.
One can see that the choices $A_{\tau, \rm in} = 0.1$ and $1$ introduce a wiggle in the radial temperature and grazing angle profiles. This artifact arises when a too large or small $\tau_{*, \rm in}$ causes an underestimated or overestimated irradiation surface height $z_s$, respectively, near the inner boundaries. With our default choice $A_{\tau, \rm in} = 0.3$ the wiggle is greatly suppressed, and the temperature profile follows a single power law $T \propto r^{-0.48}$ down to the very vicinity of the inner boundary  (see also figure~\ref{fig:JCT12_model1}).

The additional reprocessed flux $F_{\rm rep,\downarrow, in}$ significantly contributes to removing the inner boundary artifact. Without $F_{\rm rep,\downarrow, in}$, the wiggle near the inner boundary survives even with $A_{\tau, \rm in} = 0.3$ as shown in the right column of figure~\ref{fig:inner}.

Figure~\ref{fig:out} shows how $F_{\rm rep,\downarrow, out}$ works near the outer computational boundary.
Without this additional flux, the temperature profile deviates downward from the power law $T \propto r^{-0.48}$ by up to 30\% toward the outer boundary. Our prescription for $F_{\rm rep,\downarrow, out}$ reduces this deviation to $\la 1\%$. Therefore, unless the outer computational boundary represents the disk's true outer edge, one should apply this prescription to prevent temperature underestimation.

\section{Emissivity of an isothermal slab}\label{appendixC}
Here, we present an approximate model for the emissivity and absorptivity of a disk using a slab of uniform temperature $T$. 
From Kirchhoff's law for thermal radiation, the emissivity and absorptivity of this isothermal slab must be equal. This allows us to derive its emissivity by considering a special case where there is no incoming radiation to the slab.

To obtain the frequency-integrated mean intensity $J_d$ and flux $F_d$ for the slab's thermal radiation, we use the moment equations \eqref{eq:dJddtau} and \eqref{eq:dFddtau} already presented in Appedix~\ref{appendixA}, but here relax the assumption that the disk's optical thickness is infinitely large. 
Assuming that there is no incoming radiation and the outgoing radiation is isotropic at the slab's upper and lower boundaries, the radiation field must obey the boundary conditions
\beq
F_d(\tau=0) = 2\pi J_d(\tau=0),
\label{eq:bc0}
\eeq
\beq
F_d(\tau=2\tau_{\rm mid}) = -2\pi J_d(\tau=2\tau_{\rm mid}) ,
\label{eq:bc}
\eeq
where $\tau_{\rm mid}$ is the extinction optical depth to the midplane and 
$2\tau_{\rm mid}$ stands for the extinction optical thickness of the whole slab.
By solving the moment equations with the boundary conditions, we obtain 
\beq
J_d (\tau) = B_d\left[1 -\frac{\sqrt{3}\cosh(\tau_{\rm eff}-\tau_{\rm eff,mid})}{\sqrt{3}\cosh(\tau_{\rm eff,mid}) +2\sqrt{\eps} \sinh(\tau_{\rm eff,mid})}\right]
\eeq
with $\tau_{\rm eff} = \sqrt{3\eps}\tau$ and $\tau_{\rm eff,mid} = \sqrt{3\eps}\tau_{\rm mid}$.
A similar expression for the mean intensity was also derived by \citet{Miyake93},  \citet{Jang-Condell08}, \citet{Inoue09}, and \citet{Birnstiel18} but their expression is slightly different from ours because they used the two-stream approximation for the boundary condition. We have used the hemispherically isotropic outgoing boundary condition (equations~\eqref{eq:bc0} and \eqref{eq:bc}) to be consistent with the plane parallel disk models of \citet{Calvet91} and \citet{Jang-Condell04} we used in our derivation of $f_\downarrow$ (appendix~\ref{appendixA}).

The thermal flux at the upper boundary is
\beq
F_d (\tau=0) =  \frac{4\pi B_d\sqrt{\eps}\tanh(\tau_{\rm eff,mid})}{\sqrt{3} +2\sqrt{\eps} \tanh(\tau_{\rm eff,mid})}.
\eeq
The emissivity, the ratio between $F_d(\tau=0)$ and black-body radiation flux $\pi B_d$, is 
\beq
\frac{F_d (\tau=0)}{\pi B_d} = \frac{4 \sqrt{\eps}\tanh(\tau_{\rm eff,mid})}{\sqrt{3} +2\sqrt{\eps} \tanh(\tau_{\rm eff,mid})}.
\label{eq:emissivity}
\eeq
Equation~\eqref{eq:emissivity} has limiting expressions 
\beqn
\dfrac{F_d (\tau=0)}{\pi B_d} &\approx& \left\{\begin{array}{ll}
4\sqrt{\dfrac{\eps}{3}} \tau_{\rm eff,mid}, 
& \quad \tau_{\rm eff,mid} \ll 1 ,
\\[3mm]
 \dfrac{4 \sqrt{\eps}}{\sqrt{3} +2\sqrt{\eps} }, \quad
& \quad \tau_{\rm eff,mid} \gg 1 ,
 \end{array} \right.
\eeqn

\section{The thermal wave instability: radial resolution dependence}
\label{sec:Ueda21_resolution}
\begin{figure*}[t]
\begin{center}
\includegraphics[width=16.4cm, bb=0 0 648 336]{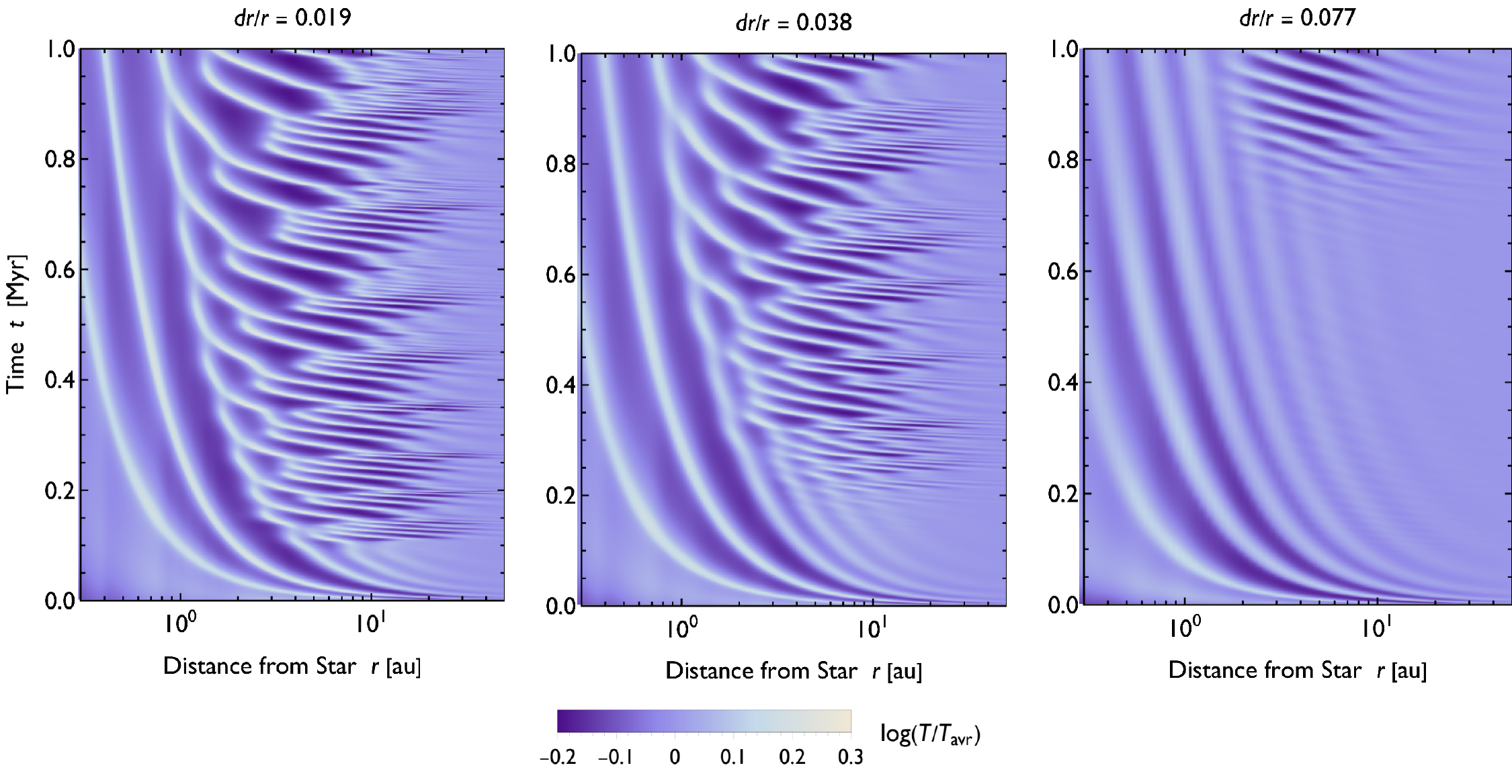}
\end{center}
\caption{
Same as figure~\ref{fig:Ueda21_wave}, but comparing the fiducial simulation with the  radial resolution of $dr/r=0.019$ (left panel, already shown in  figure~\ref{fig:Ueda21_wave}) with those with two and four times coarser resolutions $dr/r=$ 0.038 and 0.077 (center and right panels). 
}
\label{fig:Ueda21_resolution}
\end{figure*}
To examine whether the finite numerical resolution affects our simulation results for the thermal wave instability (section~\ref{sec:twi}), we performed two additional simulation runs for the same disk model but with two and four times coarser resolutions of $dr/r = 0.038$ and $dr/r = 0.077$. 

Figure~\ref{fig:Ueda21_resolution} compare the spacetime plots of $T/T_0$ from the low-resolution runs with the corresponding plot from the original simulation with 
$dr/r = 0.019$ already shown in section~\ref{sec:Ueda21_results} (figure~\ref{fig:Ueda21_wave}).
For $dr/r = 0.077$, small-scale oscillations are significantly suppressed and only become prominent after $t \ga 0.8~\rm Myr$ at $r \ga 2~\rm au$. No collisions of temperature peaks are observed, and the wave pattern is much more coherent than in higher-resolution runs. The  temperature peaks are wider than in the original run, suggesting that even the large-scale oscillations are considerably affected by the low resolution in this run. 
In the intermediate-resolution run with $dr/r = 0.038$, small-scale fluctuations become visible from $t \sim 0.2~ \rm Myr$ and the wave pattern after this time is similar to that seen in the original run. 
From this convergence study, we can conclude that our simulation is well converged at a resolution of $dr/r \la 0.04$.

\end{document}